\documentclass[10pt,letterpaper]{article}

\usepackage[T1]{fontenc}
\usepackage[utf8]{inputenc}
\usepackage[english]{babel}
\usepackage{amsmath,amssymb,amsthm}
\usepackage{XCharter}
\usepackage[charter]{newtxmath}
\IfFileExists{sourcesans.sty}{\usepackage[scale=.95]{sourcesans}}{\usepackage[scale=.95]{sourcesanspro}}
\usepackage{mathtools,bm}
\usepackage{mathrsfs}
\usepackage{graphicx,booktabs,array,tabularx,longtable,multirow}
\usepackage{xcolor,tikz}
\usepackage{colortbl}
\usetikzlibrary{arrows.meta}
\usepackage[numbers,square,sort&compress]{natbib}
\usepackage{microtype}
\usepackage{enumitem}
\usepackage{caption}
\usepackage{hyperref}
\usepackage{url}
\usepackage{placeins}
\definecolor{ink}{HTML}{183345}
\definecolor{physics}{HTML}{2B7983}
\definecolor{neural}{HTML}{7660A2}
\definecolor{feedback}{HTML}{B57A2A}
\hypersetup{colorlinks=true,linkcolor=ink,citecolor=physics,urlcolor=physics,
 pdftitle={New Wine in an Old Bottle: NeuralPOM},
 pdfauthor={Ruiqi Shu; Guangliang Liu; Ruijian Gou; Tianrun Gao; Haomin Wen; Hao Li; Xiaomeng Huang}}
\numberwithin{equation}{section}
\setlist{nosep,leftmargin=*}
\usepackage[textwidth=166mm,top=24mm,bottom=24mm,headheight=14pt,headsep=15pt,footskip=25pt]{geometry}
\usepackage{fancyhdr,titlesec}
\makeatletter
\renewcommand{\normalsize}{%
  \@setfontsize\normalsize{10.5}{14.1}%
  \abovedisplayskip=8pt plus 2pt minus 2pt
  \belowdisplayskip=8pt plus 2pt minus 2pt
  \abovedisplayshortskip=3pt plus 1pt
  \belowdisplayshortskip=5pt plus 1pt}
\renewcommand{\small}{\@setfontsize\small{9.3}{12}}
\renewcommand{\footnotesize}{\@setfontsize\footnotesize{8.5}{10.8}}
\makeatother
\normalsize
\renewenvironment{abstract}{%
 \par\begingroup\fontsize{9.7}{13.1}\selectfont\setlength{\parskip}{5pt}%
 \noindent{\sffamily\bfseries\fontsize{10}{13}\selectfont\color{physics}Abstract}\par\vspace{4pt}%
}{\par\endgroup\vspace{14pt}}
\titleformat{\section}{\sffamily\bfseries\fontsize{14.5}{18}\selectfont\color{physics}}{\thesection}{.8em}{}
\titleformat{\subsection}{\sffamily\bfseries\fontsize{11.8}{14.5}\selectfont\color{ink}}{\thesubsection}{.65em}{}
\titleformat{\subsubsection}{\sffamily\bfseries\normalsize\color{ink}}{\thesubsubsection}{.65em}{}
\titlespacing*{\section}{0pt}{19pt plus 3pt minus 2pt}{8pt}
\titlespacing*{\subsection}{0pt}{14pt plus 2pt minus 2pt}{6pt}
\titlespacing*{\subsubsection}{0pt}{10pt plus 2pt}{5pt}
\DeclareCaptionFont{reportcaption}{\fontsize{9.2}{12.2}\selectfont}
\fancypagestyle{plain}{\fancyhf{}\fancyfoot[R]{\sffamily\fontsize{8.5}{10}\selectfont\color{physics}\thepage}}
\newcommand{\AuthorList}{%
Ruiqi Shu\textsuperscript{1,2,\ensuremath{\dagger}},
Guangliang Liu\textsuperscript{1,3,4,\ensuremath{\dagger}},
Ruijian Gou\textsuperscript{5,\ensuremath{\dagger}},
Tianrun Gao\textsuperscript{6},
Haomin Wen\textsuperscript{7},
Hao Li\textsuperscript{8,7,2,*},
Xiaomeng Huang\textsuperscript{1,*}}

\newcommand{\AffiliationList}{%
\textsuperscript{1}Department of Earth System Science, Tsinghua University, Beijing, China\\
\textsuperscript{2}Shanghai Academy of Artificial Intelligence for Science, Shanghai, China\\
\textsuperscript{3}State Key Laboratory of Physical Oceanography, Institute of Oceanographic Instrumentation, Shandong Academy of Sciences, Qingdao, China\\
\textsuperscript{4}Key Laboratory of Computing Power Network and Information Security, Ministry of Education, Shandong Computer Science Center, Qilu University of Technology (Shandong Academy of Sciences), Jinan, China\\
\textsuperscript{5}Department of Ocean Big Data and Artificial Intelligence, Laoshan National Laboratory, Qingdao, China\\
\textsuperscript{6}School of Computing and Data Science, The University of Hong Kong\\
\textsuperscript{7}Shanghai Innovation Institute, Shanghai, China\\
\textsuperscript{8}Artificial Intelligence Innovation and Incubation Institute, Fudan University, Shanghai, China}

\newcommand{\AuthorNotes}{%
\textsuperscript{\ensuremath{\dagger}}These authors contributed equally.\\
\textsuperscript{*}Correspondence: Hao Li (\href{mailto:lihao_lh@fudan.edu.cn}{lihao\_lh@fudan.edu.cn}) and Xiaomeng Huang (\href{mailto:hxm@tsinghua.edu.cn}{hxm@tsinghua.edu.cn}).}

\newcommand{\MainMethodFigure}{%
\begin{figure*}[t]
\centering
\includegraphics[width=\textwidth]{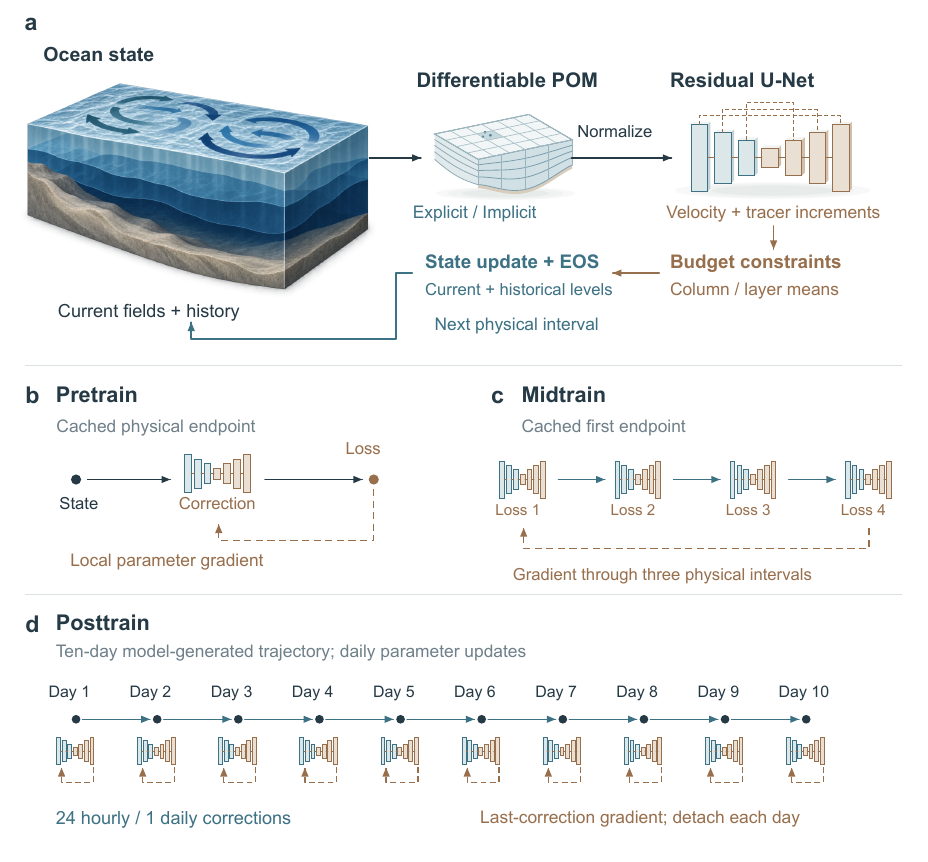}
\caption{\textbf{Physical--neural coupling and three-stage training.} \textbf{a,} At each coupling step, the physical dynamical framework supplies the ocean state to a residual U-Net, which returns conservation-constrained increments to update the physical fields. \textbf{b,} Pretrain optimizes single-step corrections on cached physical forecasts. \textbf{c,} Midtrain unrolls four correction steps across three online differentiable physical integration intervals. \textbf{d,} Posttrain integrates 10-day closed-loop trajectories under high-frequency (\(\Delta=1\,\mathrm h\)) or low-frequency (\(\Delta=1\,\mathrm{day}\)) coupling with daily parameter updates and inter-day gradient detachment. Warm brown dashed arrows denote active parameter-gradient paths.}
\label{fig:method}
\end{figure*}}

\makeatletter
\newcommand{\EditorialTOC}{%
  \par\vspace{6pt}%
  \noindent\begin{minipage}{\textwidth}
  {\color{physics}\hrule height .55pt}%
  \vspace{8pt}%
  {\noindent\sffamily\bfseries\fontsize{10.5}{13}\selectfont\color{physics}Contents\par}%
  \vspace{3pt}%
  \begingroup
  \setlength{\parskip}{0.8pt}%
  \renewcommand*\l@section[2]{%
    \ifnum \c@tocdepth >\z@
      \addvspace{3.2pt}%
      \setlength\@tempdima{1.5em}%
      \begingroup
        \parindent \z@ \rightskip \@pnumwidth
        \parfillskip -\@pnumwidth
        \leavevmode \sffamily\bfseries\fontsize{9.1}{11.4}\selectfont\color{physics}%
        \advance\leftskip\@tempdima
        \hskip -\leftskip
        ##1\nobreak\leaders\hbox{$\m@th\mkern 4.5mu\textcolor{physics!40}{\cdot}\mkern 4.5mu$}\hfill
        \nobreak\hb@xt@\@pnumwidth{\hss\sffamily\bfseries\fontsize{9.0}{11.4}\selectfont\color{physics}##2}\par
      \endgroup
    \fi}%
  \renewcommand*\l@subsection[2]{%
    \ifnum \c@tocdepth >\@ne
      \setlength\@tempdima{2.1em}%
      \begingroup
        \parindent 1.5em \rightskip \@pnumwidth
        \parfillskip -\@pnumwidth
        \leavevmode \fontsize{8.6}{10.8}\selectfont\color{ink}%
        \advance\leftskip 3.6em
        \hskip -2.1em
        ##1\nobreak\leaders\hbox{$\m@th\mkern 4.5mu\textcolor{ink!30}{\cdot}\mkern 4.5mu$}\hfill
        \nobreak\hb@xt@\@pnumwidth{\hss\sffamily\fontsize{8.5}{10.8}\selectfont\color{ink}##2}\par
      \endgroup
    \fi}%
  \@starttoc{toc}%
  \endgroup
  \vspace{8pt}%
  {\color{physics}\hrule height .55pt}%
  \end{minipage}\par\vspace{10pt}%
}
\makeatother

\begin{document}
\thispagestyle{plain}
{\fontsize{9}{11}\selectfont\color{physics}\textsc{NeuralPOM}\par}
\vspace{19pt}
{\fontsize{35}{39}\selectfont\color{ink}New Wine in an Old Bottle\par}
\vspace{9pt}
{\fontsize{21}{26}\selectfont\raggedright\color{ink}NeuralPOM: A Hybrid Ocean Forecast Model\\with Differentiable Dynamics and Staged Learning\par}
\vspace{19pt}
{\fontsize{11}{14}\selectfont\raggedright\AuthorList\par}
\vspace{7pt}
{\fontsize{8.8}{11.4}\selectfont\raggedright\AffiliationList\par}
\vspace{5pt}
{\fontsize{8.3}{11}\selectfont\raggedright\AuthorNotes\par}
\vspace{18pt}
{\color{physics}\hrule height .65pt}
\vspace{19pt}

\begin{abstract}
Coarse-grid ocean circulation models rely on empirical subgrid parameterizations that struggle to capture mesoscale eddies, whereas purely data-driven neural forecasters accumulate trajectory drift, violate tracer conservation, and damp small-scale variability over extended integrations. Here we present NeuralPOM, a hybrid ocean forecast model that couples differentiable explicit mode-split and semi-implicit free-surface implementations of the Princeton Ocean Model (POM) to a conservation-constrained neural corrector. The three-dimensional primitive-equation solver advances the ocean state, while the neural network periodically updates the physical fields at high-frequency (\(1\,\mathrm h\)) or low-frequency (\(1\,\mathrm{day}\)) intervals under volume, heat, and salt conservation constraints. Using a fine-grid reference dataset exceeding 100,000 model days in a wind-driven double-gyre basin, we introduce a three-stage training procedure comprising offline single-step pretraining, multi-step differentiable training through the ocean solver, and multi-day closed-loop rollout training. Over 100-day forecasts, NeuralPOM reduces errors across all prognostic fields relative to coarse-grid physical closures and direct neural forecasters, preserving mesoscale kinetic energy while improving eddy phase alignment. Controlled ablations show that closed-loop training suppresses multi-day drift, whereas backpropagating gradients through the differentiable solver stabilizes high-frequency coupling. Finally, because high-frequency tracer updates can introduce density anomalies via the equation of state that excite spurious baroclinic pressure gradients, we introduce Density-Normal Damping (DND) to attenuate the density-modifying (diapycnal) tracer component while retaining the density-compensated (isopycnal) correction.
\end{abstract}

\EditorialTOC
\section{Introduction}
\label{sec:introduction}

Forecasting the physical state of the ocean requires predicting the coupled three-dimensional evolution of horizontal and vertical currents, free-surface elevation, potential temperature, salinity, seawater density, and turbulent mixing. Operational and climate-scale ocean models address this problem by numerically integrating the hydrostatic, Boussinesq primitive equations under prescribed atmospheric forcing and basin geometry \cite{FoxKemper2019OceanModels,Heimbach2025OceanForecasting}. At grid spacings affordable for extended ensembles or multi-month integrations, however, numerical models only partially resolve the baroclinic instabilities and mesoscale eddies that govern much of the ocean's kinetic energy and tracer transport \cite{Hewitt2020Mesoscale}, while empirical subgrid closures must balance numerical stability against excessive dissipation of resolved scales. Conversely, purely data-driven neural forecasters can advance daily ocean states at lower computational cost than numerical solvers \cite{Cui2025EddyingOcean,ElAouni2025GLONET,Holmberg2025SeaCast,Niu2026FuXiOcean}, yet over multi-week autoregressive integrations they exhibit compounding trajectory drift \cite{Chattopadhyay2024OceanNet,Cui2025EddyingOcean,ElAouni2025GLONET}, progressive attenuation of near-grid mesoscale variability under pointwise mean-squared-error objectives \cite{Leith1974MonteCarlo,Kochkov2024NeuralGCM,Chattopadhyay2024OceanNet}, and lack built-in volume, heat, and salt conservation constraints across interior layers \cite{Kochkov2024NeuralGCM,Zanna2025HybridOceanFramework}.

A natural question is whether numerical circulation models and machine learning can be combined into a hybrid framework that retains the conservation laws and dynamical balance of a numerical solver while leveraging data-driven corrections for unresolved processes. Recent studies have embedded learned subgrid parameterizations within idealized and regional ocean models \cite{BoltonZanna2019,ZannaBolton2020,GuillauminZanna2021,Ross2023_8,Perezhogin2024Stable,Kamm2026Atlantic}, demonstrated end-to-end differentiable training in atmospheric general circulation models \cite{Kochkov2024NeuralGCM}, and coupled physics-guided neural correctors and assimilation modules with ocean dynamical frameworks \cite{Shu2025PhysicsGuidedMHW,Zanna2025HybridOceanFramework,Shu2026OceanE2E,Shu2026HybridOM}. Most ocean hybrid studies to date, however, have focused on idealized quasi-geostrophic, barotropic, or shallow-water configurations, leaving open how neural state updates interact with a complete three-dimensional, terrain-following primitive-equation circulation model that prognostically evolves barotropic and baroclinic velocity modes, a free surface, active temperature and salinity, a nonlinear equation of state (EOS), and a two-equation vertical turbulence closure.

Coupling a neural corrector to a three-dimensional primitive-equation ocean model introduces three physical and numerical challenges that do not arise in simplified models. First, because each correction alters the state trajectory visited by the physical dynamical framework, a network trained only on offline single-step targets experiences distribution shift during closed-loop integration \cite{Ross2023_8,Ouala2024OnlineCalibration}, requiring staged training procedures that combine multi-step differentiable training through the dynamical framework with closed-loop rollout training \cite{Frezat2022_10,List2025,Yan2025AdjointOnline}. Second, advancing the solver across time steps requires complete multi-time-level state histories, including current and backward leapfrog levels, barotropic--baroclinic mode-coupling transports, and turbulent kinetic energy and mixing-length fields; unconstrained state updates that violate leapfrog time-level consistency or column volume budgets excite computational modes and fast external gravity waves. Third, potential temperature and salinity are active thermodynamic variables that enter the horizontal momentum equations through the nonlinear EOS \cite{Mellor1991EOS} and hydrostatic baroclinic pressure-gradient force, while modulating vertical viscosity and diffusivity through the buoyancy frequency in the turbulence closure \cite{MellorYamada1982}. Consequently, a tracer increment that reduces instantaneous pointwise error at a single step can still introduce small-scale density anomalies, generate spurious baroclinic pressure-gradient tendencies, and perturb vertical mixing over subsequent integration steps.

Here we present NeuralPOM, a hybrid ocean forecasting system built upon the three-dimensional Princeton Ocean Model (POM) \cite{BlumbergMellor1987POM,Xu2015gpuPOM}. NeuralPOM combines two core components: differentiable PyTorch POM dynamical frameworks---covering both explicit mode-split and semi-implicit free-surface formulations verified against double-precision Fortran POM across idealized and coastal benchmarks---and a residual U-Net corrector that periodically updates horizontal velocity, potential temperature, and salinity under volume, heat, and salt conservation constraints. Using a fine-grid reference ensemble exceeding 100,000 model days in a wind-driven double-gyre ocean basin, we train NeuralPOM through a three-stage training procedure comprising offline single-step pretraining, multi-step differentiable training through the physical dynamical framework, and 10-day closed-loop rollout training. Over 100-day forecasts, NeuralPOM reduces forecast errors across all prognostic variables relative to coarse-grid physical parameterizations and direct neural forecasting architectures \cite{Wu2025TritonV1,Gao2022SimVP,Shi2015ConvLSTM,Wu2023PastNet}. Stage progressions and controlled route ablations demonstrate that multi-step differentiable training and closed-loop rollout training play complementary roles: backpropagating gradients through the differentiable dynamical framework during multi-step training stabilizes thermohaline and circulation evolution under high-frequency hourly coupling, while multi-day closed-loop rollout training suppresses cumulative trajectory drift across both low-frequency daily and high-frequency hourly coupling regimes. Furthermore, diagnostic perturbation experiments and 100-day closed-loop integrations reveal that under high-frequency hourly coupling, uncoordinated temperature and salinity increments project onto density anomalies via the nonlinear EOS and trigger spurious baroclinic pressure-gradient feedback; decomposing each layerwise tracer increment relative to the local EOS density gradient and applying Density-Normal Damping (DND) to attenuate the density-modifying (diapycnal) component while preserving the density-compensated (isopycnal) water-mass correction suppresses this instability and consistently improves long-range skill. Finally, wavenumber spectral analysis shows that NeuralPOM maintains physical eddy kinetic energy and near-grid variability while reducing spectral error through improved mesoscale phase alignment.

\section{The NeuralPOM Framework}
\label{sec:methods}

NeuralPOM alternates two operations across forecast cycles: a differentiable three-dimensional primitive-equation dynamical framework advances the complete numerical ocean state over a physical integration interval \(\Delta\), and a conservation-constrained neural corrector updates the physical fields at the end of each interval (Fig.~\ref{fig:method}). We investigate both high-frequency (\(\Delta = 1\,\mathrm h\)) and low-frequency (\(\Delta = 1\,\mathrm{day}\)) coupling regimes to examine how the frequency of neural state updates interacts with three-dimensional primitive-equation dynamics.

\MainMethodFigure

\subsection{Differentiable primitive-equation frameworks}
\label{sec:methods-physics}

Both dynamical frameworks integrate the hydrostatic, Boussinesq primitive equations of the Princeton Ocean Model (POM) on a staggered Arakawa-C horizontal grid with terrain-following \(\sigma\)-coordinates \cite{BlumbergMellor1987POM}. With horizontal velocity \(\boldsymbol u_h=(u,v)\), vertical velocity \(w\), material derivative \(D_t\), Coriolis parameter \(f\), reference density \(\rho_0\), hydrostatic pressure \(p\), in-situ density \(\rho\), and prognostic tracers \(C\in\{T,S\}\), the continuous governing equations are
\begin{align}
 D_t\boldsymbol u_h+f\boldsymbol e_z\times\boldsymbol u_h
 &=-\rho_0^{-1}\nabla_h p+\mathcal D_m,
 &\partial_zp&=-g\rho,\label{eq:method-momentum}\\
 \nabla_h\cdot\boldsymbol u_h+\partial_zw&=0,
 &D_tC&=\mathcal D_C,
 \label{eq:method-tracers}
\end{align}
where \(\mathcal D_m\) and \(\mathcal D_C\) represent subgrid momentum and tracer mixing. Seawater density \(\rho(T,S,P)\) is evaluated from the nonlinear Mellor equation of state (EOS) \cite{Mellor1991EOS}, vertical viscosity \(K_M\) and diffusivity \(K_H\) are obtained from the prognostic two-equation Mellor--Yamada level-2.5 turbulence closure for turbulent kinetic energy and length scale \((q^2,q^2l)\) \cite{MellorYamada1982}, and reference horizontal dissipation follows the harmonic Smagorinsky formulation (\(A_H = 0.2A_M\)).

The free-surface elevation \(\eta\) evolves according to the vertically integrated continuity equation \(\partial_t\eta+\nabla_h\cdot(H\boldsymbol U)=0\), where \(H=h+\eta\) is the total water-column depth and \(\boldsymbol U=(U,V)\) is the depth-averaged horizontal velocity. We implemented two differentiable PyTorch dynamical frameworks that share identical three-dimensional baroclinic, thermohaline, and turbulence operators at the internal time step \(\Delta t = 600\,\mathrm s\) but differ in their numerical integration of fast external gravity waves. The explicit mode-split framework (\emph{Explicit}) directly follows the standard split-explicit formulation of the original Fortran POM \cite{BlumbergMellor1987POM}: within each internal step, the depth-integrated barotropic mode is subcycled explicitly over \(N_e=30\) leapfrog steps (\(\delta t = 20\,\mathrm s\)) with Robert--Asselin time filtering, accumulating interval-averaged free-surface elevations and face transports to maintain volume consistency with the three-dimensional baroclinic mode.

In the semi-implicit free-surface framework (\emph{Implicit}), external subcycling is replaced by an elliptic Helmholtz solve for the forward free-surface elevation \(\eta^f\) at \(\Delta t = 600\,\mathrm s\) using preconditioned conjugate gradients (PCG) with a discrete sine transform preconditioner; during reverse-mode differentiation, a custom implicit-differentiation adjoint solves the area-weighted self-adjoint linear system directly rather than backpropagating through the PCG iterations, accounting for the sensitivity of the elliptic operator to total water depth. Appendix~\ref{app:physics} presents the discrete governing equations and adjoint formulation, and Appendix~\ref{app:validation} verifies both PyTorch frameworks against double-precision Fortran POM across double-gyre, seamount, idealized coastal, and realistic Jiaozhou Bay tidal benchmarks. In both formulations, potential temperature and salinity force the momentum and turbulence equations through the diagnostic density \(\rho\) via hydrostatic baroclinic pressure gradients and vertical buoyancy stratification, and advancing the solver across intervals requires the complete 45-variable state vector across current (\(n\)) and backward (\(b\)) leapfrog time levels.

\subsection{Inference framework: Constrained state updates and density-normal damping}
\label{sec:methods-correction}

With the differentiable primitive-equation solver in place, coupled forecast inference proceeds by alternating physical integration with periodic neural state updates at either high frequency (\(\Delta=1\,\mathrm h\)) or low frequency (\(\Delta=1\,\mathrm{day}\)). Let \(x_n^+\) denote the complete numerical ocean state after correction cycle \(n\), \(\mathcal P^\Delta\) the physical integration operator over interval \(\Delta\), and \(\mathcal K_\theta\) the constrained neural correction operator. Each coupled forecast cycle advances according to
\begin{equation}
 x_{n+1}^-=\mathcal P^\Delta(x_n^+),\qquad
 x_{n+1}^+=\mathcal K_\theta(x_{n+1}^-).
 \label{eq:method-coupling}
\end{equation}
At the end of each physical interval, a residual U-Net takes 101 standardized input channels---comprising \((u^n,u^b,v^n,v^b,T^n,S^n)\) together with the external-mode fields \(\eta_{\mathrm E}^n\) and \((U^n,U^b,V^n,V^b)\)---and predicts 60 raw increment channels \((\delta u,\delta v,\delta T,\delta S)\) across the 15 active vertical \(\sigma\)-layers (Appendix~\ref{app:training-network}).

To ensure that the neural increments respect the conservation laws and time-stepping structure of the primitive-equation solver, raw network outputs are filtered through a sequence of physical projections before updating the model state (Appendix~\ref{app:training-correction}). Horizontal velocity increments are first projected to zero thickness-weighted column mean (\(\sum_{k=0}^{14}\Delta\sigma_k\delta u_{ijk}=0\) and \(\sum_{k=0}^{14}\Delta\sigma_k\delta v_{ijk}=0\)), restricting direct momentum updates to the vertical baroclinic shear so that depth-integrated volume transport remains governed by the external continuity equation. Temperature and salinity increments are similarly projected to zero volume-weighted horizontal mean on every active \(\sigma\)-layer (\(\sum_{ij}w_{ijk}\delta T_{ijk}=0\) and \(\sum_{ij}w_{ijk}\delta S_{ijk}=0\), where \(w_{ijk}=A_{ij}m_{ij}H_{\mathrm B,ij}\Delta\sigma_k\) is the wet cell volume), redistributing heat and salt laterally without introducing net layer sources or sinks.

Within each \(\sigma\)-layer, the standardized tracer increment \(z=(\delta T'/s_T,\delta S'/s_S)\) is further decomposed relative to the local EOS density gradient into two orthogonal, layer-mean-preserving components: a density-modifying (diapycnal) component \(Ez\) aligned with the local density normal, and a density-compensated (isopycnal) component \((I-E)z\) tangent to local isopycnals. Scaling the density-modifying component by \(c\in[0,1]\) yields the filtered tracer update
\begin{equation}
 z_c=(I-E)z+cEz.
 \label{eq:method-dnd}
\end{equation}
Setting \(c=1\) gives the \emph{Undamped} tracer update, whereas setting \(c=3/4\) defines \emph{Density-Normal Damping} (DND), which attenuates the density-modifying component by \(25\%\) in amplitude (\(43.75\%\) in squared norm) while fully preserving the density-compensated water-mass correction (Section~\ref{sec:analysis} and Appendix~\ref{app:theory}). To prevent extreme local corrections without violating the zero-mean conservation constraints, the standardized velocity and tracer increments are proportionally rescaled whenever their maximum magnitude exceeds \(z_{\max}=5\). The resulting physical increments are then applied synchronously to both the current (\(n\)) and backward (\(b\)) leapfrog time levels (\(f^n\leftarrow f^n+\widehat{\delta f}\) and \(f^b\leftarrow f^b+\widehat{\delta f}\) for \(f\in\{u,v,T,S\}\)), which shifts the model state without altering the leapfrog difference \(f^n-f^b\) or exciting computational modes. Finally, three-dimensional density \(\rho\) is immediately updated from the corrected \((T,S)\) fields via the nonlinear EOS, while the remaining prognostic and diagnostic fields (\(\eta\), \(\Omega\), turbulence variables, and barotropic transports) adjust dynamically during the next physical integration \(\mathcal P^\Delta\).

\subsection{Training framework: Physics-informed objective and three-stage training procedure}
\label{sec:methods-objective}
\label{sec:methods-stages}

While the constrained inference operator \(\mathcal K_\theta\) enforces instantaneous conservation and leapfrog consistency at each coupling step, maintaining accuracy and dynamical stability over multi-week rollouts requires a physics-informed objective paired with a staged training procedure. At each scored correction step, the updated coarse-grid state \(x\) is evaluated against the coarse-grained fine-grid reference state \(y\) using a composite loss \(\ell(x,y)\) (Appendix~\ref{app:training-loss}). In addition to volume-weighted mean squared errors in horizontal velocity, potential temperature, and salinity, \(\ell(x,y)\) incorporates a hydrostatic baroclinic pressure-gradient consistency term, a horizontal frontal-gradient term, and (during multi-step training) a sea-surface height term to penalize dynamically imbalanced corrections.

Training a randomly initialized corrector directly on long closed-loop rollouts is computationally expensive and prone to early trajectory divergence, whereas a network trained only on offline single-step targets suffers from distribution shift and unmitigated dynamical feedbacks once coupled in a closed loop. We therefore train \(\mathcal K_\theta\) sequentially through three stages---\emph{Pretrain}, \emph{Midtrain}, and \emph{Posttrain}---that progressively extend the coupling horizon from single-step offline states to multi-step differentiable physics and multi-day closed-loop rollouts (Fig.~\ref{fig:method}b--d and Appendix~\ref{app:training-stages}).

In the first stage, single-step pretraining (\emph{Pretrain}), the network learns an initial correction mapping without differentiating through the ocean solver. Starting from a reference state \(x_0\), the physical framework advances one interval offline to produce a cached pre-correction forecast \(x_1^-=\mathcal P^\Delta(x_0)\), and \(\mathcal K_\theta\) is optimized by minimizing the single-step loss against the reference target \(y_1\):
\begin{equation}
 J_{\mathrm{pre}}(\theta)=\mathbb E\bigl[\ell\bigl(\mathcal K_\theta(x_1^-),\,y_1\bigr)\bigr],\qquad
 x_1^-=\mathcal P^\Delta(x_0),
 \label{eq:method-pre}
\end{equation}
with \(\mathcal P^\Delta\) outside the automatic-differentiation graph.

Because single-step pretraining cannot observe how a state update alters subsequent physical evolution, the second stage, multi-step differentiable training (\emph{Midtrain}), teaches the corrector to account for the dynamical response of the primitive-equation solver. Initialized from the Pretrain checkpoint, the coupled model is unrolled across four correction cycles connected by three online physical integration intervals retained within the differentiable solver graph (spanning \(4\,\mathrm h\) for high-frequency \(\Delta=1\,\mathrm h\) coupling and \(4\,\mathrm{days}\) for low-frequency \(\Delta=1\,\mathrm{day}\) coupling):
\begin{equation}
 J_{\mathrm{mid}}(\theta)=\frac14\sum_{j=1}^{4}\ell(x_j^+,y_j),\qquad
 x_j^+=\mathcal K_\theta\bigl(\mathcal P^\Delta(x_{j-1}^+)\bigr).
 \label{eq:method-mid}
\end{equation}
Backpropagating through \(\mathcal P^\Delta\) allows endpoint errors at steps \(j>1\)---including free-surface elevation and velocity errors---to constrain earlier state updates at step \(j-1\) through the linearized primitive equations.

Even after four-step differentiable training, extended autoregressive integrations encounter cumulative trajectory drift over multi-day horizons where full backpropagation through time is memory-prohibitive. In the final stage, closed-loop rollout training (\emph{Posttrain}), the model is initialized from the Midtrain checkpoint and integrated continuously over 10-day closed-loop rollouts (\(d=1,\dots,10\)) to adapt \(\mathcal K_\theta\) to the closed-loop state distribution. Let \(K=\mathrm{1\,day}/\Delta\) denote the number of coupling cycles per day (\(K=1\) for low-frequency \(\Delta=1\,\mathrm{day}\) coupling and \(K=24\) for high-frequency \(\Delta=1\,\mathrm h\) coupling). On each rollout day \(d\), starting from the previous day's detached state \(x_{d,0}^+=\operatorname{sg}(x_{d-1,K}^+)\), the coupled system advances through \(K\) cycles with the stop-gradient operator \(\operatorname{sg}(\cdot)\) detaching the physical state before each neural update:
\begin{equation}
 x_{d,k}^-=\mathcal P^\Delta(x_{d,k-1}^+),\qquad
 x_{d,k}^+=\mathcal K_\theta\bigl(\operatorname{sg}(x_{d,k}^-)\bigr)\qquad (k=1,\dots,K).
 \label{eq:method-post-rollout}
\end{equation}
Network parameters \(\theta\) are updated once at the end of each day (\(k=K\)) against the daily reference state \(y_d\) by minimizing
\begin{equation}
 J_{\mathrm{post},d}(\theta)=\ell(x_{d,K}^+,y_d),
 \label{eq:method-post}
\end{equation}
before detaching \(x_{d,K}^+\) to initialize day \(d+1\). Under this formulation, the first \(K-1\) cycles of each day evolve the closed-loop trajectory in forward-only mode and only the final cycle of each day tracks parameter gradients, suppressing multi-day trajectory drift without backpropagating across 10-day horizons.

\section{Double-Gyre Benchmark and Experimental Protocol}
\label{sec:experimental-design}

\subsection{Wind-driven double-gyre reference dataset}
\label{sec:methods-data}
\label{sec:design-data}

\begin{figure}[!htbp]
\centering
\includegraphics[width=0.82\textwidth]{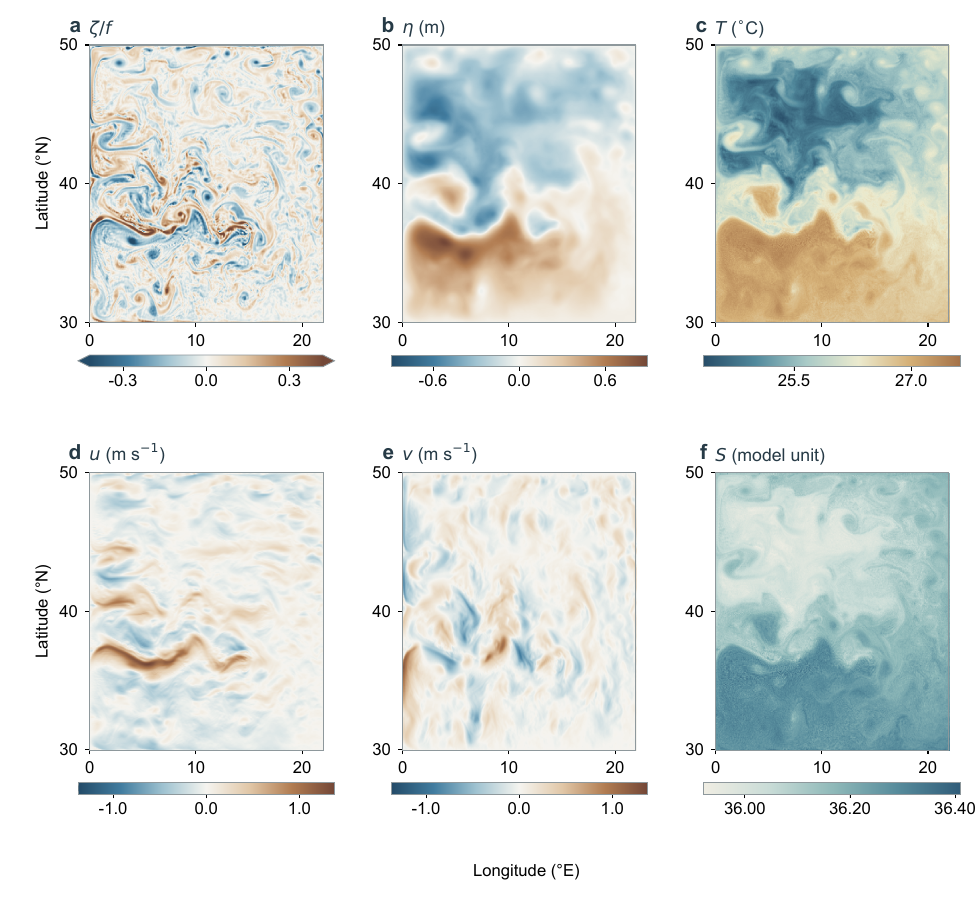}
\caption{\textbf{Developed three-dimensional circulation in the fine-grid wind-driven double-gyre reference dataset.} Instantaneous uppermost-layer normalized relative vorticity \(\zeta/f\), free-surface elevation, potential temperature, zonal velocity, meridional velocity, and salinity from the fine-grid Explicit reference simulation (model day 10,298).}
\label{fig:developed-fields}
\end{figure}

We evaluate NeuralPOM in an idealized three-dimensional, wind-driven double-gyre ocean basin spanning \(30^\circ\mathrm N\) to \(50^\circ\mathrm N\) across a \(2500\,\mathrm{km}\) domain with a flat \(2000\,\mathrm m\) bottom and 15 active terrain-following \(\sigma\)-layers (Fig.~\ref{fig:developed-fields}). Steady sinusoidal zonal wind stress forces a subtropical and subpolar gyre pair separated by an unstable eastward jet and an energetic mesoscale eddy field. Potential temperature is initialized from an exponential vertical stratification profile and salinity from a uniform value, with zero surface heat and freshwater fluxes.

For each dynamical framework (\emph{Explicit} and \emph{Implicit}), we generated a multi-member fine-grid reference ensemble at eddy-resolving resolution (\(\sim 7\,\mathrm{km}\) horizontal spacing, \(354\times 322\times 16\) grid) exceeding 100,000 model days in total, and conservatively restricted the trajectories onto an eddy-permitting coarse forecast grid (\(\sim 28\,\mathrm{km}\) spacing, \(90\times 82\times 16\) grid) to provide both low-frequency daily and high-frequency hourly reference targets (Appendix~\ref{app:training-data}). The ensemble is partitioned into independent training, validation, and test splits; all out-of-sample evaluations are initialized from 16 evenly spaced test origins on an unseen ensemble member and integrated continuously for 100 days.

\subsection{Baselines and evaluation protocol}
\label{sec:design-evaluation}
\label{sec:design-policies}
\label{sec:design-mechanism}

Our experimental design crosses the two dynamical frameworks (\emph{Implicit} and \emph{Explicit}), two coupling frequencies---low-frequency daily (\(\Delta=1\,\mathrm{day}\)) and high-frequency hourly (\(\Delta=1\,\mathrm h\))---and the three training stages (\emph{Pretrain}, \emph{Midtrain}, and \emph{Posttrain}; Appendix~\ref{app:training-optimization}). Forecast skill is quantified over 100-day integrations across all 16 test origins using volume-weighted root-mean-square errors (RMSE) and radially averaged horizontal velocity wavenumber spectra (Appendix~\ref{app:evaluation-metrics}).

We benchmark NeuralPOM against two categories of baselines. First, on the uncorrected coarse-grid POM solver, we compare three classical horizontal subgrid closures (Appendix~\ref{app:evaluation-baselines}): standard second-order harmonic \emph{Smagorinsky} diffusion (the primary physical baseline), scale-selective fourth-order \emph{Biharmonic} viscosity, and a kinetic-energy \emph{Backscatter} parameterization that recycles a fraction of biharmonically dissipated energy into the resolved flow. Second, we compare against five direct spatiotemporal neural forecasters---\emph{Triton} \cite{Wu2025TritonV1}, \emph{ResUNet}, \emph{SimVP} \cite{Gao2022SimVP}, \emph{ConvLSTM} \cite{Shi2015ConvLSTM}, and \emph{PastNet} \cite{Wu2023PastNet}---which autoregressively advance the full daily ocean state without a numerical solver and are trained through both single-step pretraining (\emph{Pre}) and 10-day closed-loop rollout training (\emph{Post}) (Appendix~\ref{app:ai-comparison}).

To isolate the mechanisms governing hybrid forecast stability, we conduct three sets of controlled ablations. First, to evaluate directional tracer damping at inference, networks trained with \(c=1\) in Eq.~\eqref{eq:method-dnd} are deployed with frozen weights under three tracer update rules: \emph{Undamped} (\(c=1\)), \emph{Density-Normal Damping} (\emph{DND}, \(c=3/4\)), and an isotropic norm-matched control (\emph{Uniform}) that scales the entire tracer increment uniformly by a scalar factor \(\kappa\) calibrated to match the mean increment norm of DND (Appendix~\ref{app:theory-implementation}). Second, to separate the effect of backpropagating gradients through the differentiable solver during \emph{Midtrain} from multi-step rollout exposure and closed-loop \emph{Posttrain}, we compare three training routes at matched update budgets (Appendix~\ref{app:training-routes}): the full three-stage procedure (\emph{Full}, \(\mathrm{Pre\to Mid\to Post}\)), skipping \emph{Midtrain} (\emph{Skip}, \(\mathrm{Pre\to Post}\)), and a detached-gradient control (\emph{Detached}, \(\mathrm{Pre\to Mid_{detached}\to Post}\)) that executes the identical 4-step \emph{Midtrain} rollouts while detaching state gradients before each physical interval \(\mathcal P^\Delta\). Third, to trace how a single tracer correction propagates through the primitive equations without subsequent neural intervention, we sample 36 pre-correction states from the hybrid forecast trajectories, apply the neural corrector once at \(t=0\)---using an identical velocity update paired with four tracer update rules (\emph{no-TS}, \emph{Undamped}, \emph{Uniform}, and \emph{DND})---and then turn off the neural corrector to integrate the pure physical solver forward for 72 hours (Appendix~\ref{app:theory-implementation}).

\section{Forecast Results}
\label{sec:results}

\subsection{Whole-state forecast skill across dynamical frameworks}
\label{sec:results-forecasts}

\begin{figure}[tbp]
\centering
\includegraphics[width=\linewidth]{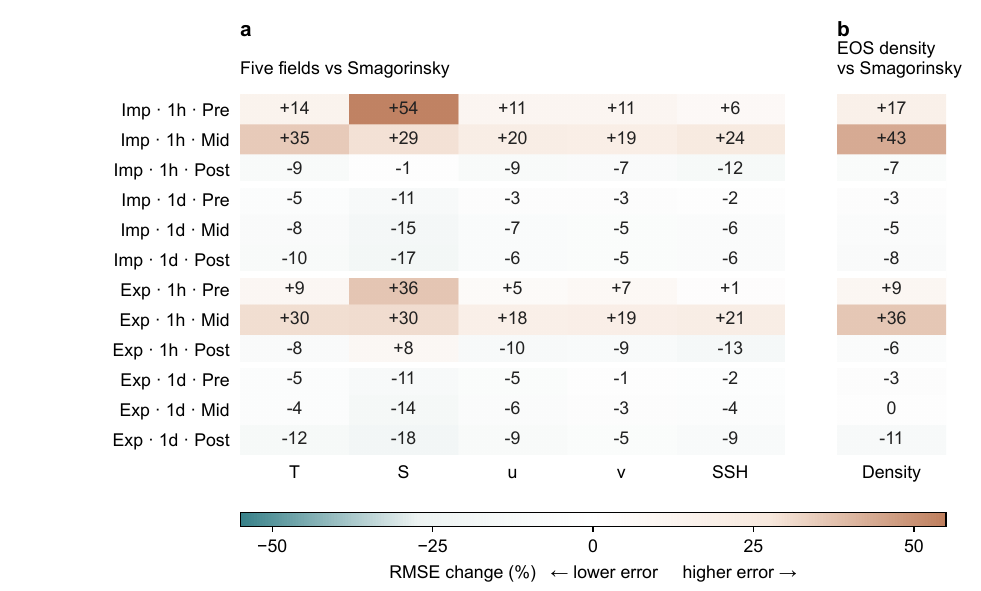}
\caption{\textbf{Forecast skill across training stages and correction intervals.} \textbf{a,} Percentage change in 100-day RMSE relative to the Smagorinsky physical baseline for all twelve hybrid configurations deployed with DND. \textbf{b,} Corresponding percentage change in 100-day EOS density RMSE. Negative values indicate improvements over Smagorinsky; annotations are rounded to whole percentages. Imp and Exp denote the semi-implicit and explicit frameworks.}
\label{fig:forecast-skill}
\end{figure}

Across 100-day forecasts initialized from the 16 out-of-sample test origins, NeuralPOM improves whole-state forecast skill relative to the uncorrected coarse-grid Smagorinsky baseline in both the semi-implicit (Implicit) and explicit mode-split (Explicit) dynamical frameworks (Table~\ref{tab:skill} and Fig.~\ref{fig:forecast-skill}). Under low-frequency daily coupling (\(\Delta=1\,\mathrm{day}\)), both DND and Undamped deployments lower 100-day root-mean-square error (RMSE) across all five prognostic variables, reducing horizontal velocity and sea-surface height (SSH) errors by \(5\%\text{--}10\%\), potential temperature error by \(10\%\text{--}12\%\), salinity error by \(15\%\text{--}18\%\), and diagnostic equation-of-state (EOS) density error by \(8\%\text{--}11\%\) (Fig.~\ref{fig:forecast-skill}b). Even without direct neural updates to \(\eta\), SSH errors decrease by \(6\%\text{--}9\%\) because updates to baroclinic velocity shear and interior buoyancy stratification adjust the sea surface through barotropic--baroclinic mode coupling during the subsequent physical integration.

\begin{table}[tbp]
\centering
\caption{\textbf{Whole-state 100-day forecast skill relative to the coarse-grid Smagorinsky baseline.} Percentage change in pooled 100-day RMSE across the 16 test origins relative to the uncorrected Smagorinsky physical model of the same dynamical framework (Implicit or Explicit); negative values indicate error reductions. Hybrid rows report the final Posttrain checkpoints deployed with Density-Normal Damping (DND, \(c=3/4\)) and the undamped tracer update (Undamped, \(c=1\)). Seawater density \(\rho\) is evaluated diagnostically via the nonlinear POM equation of state.}
\label{tab:skill}
\small
\setlength{\tabcolsep}{4.5pt}
\begin{tabular}{lllrrrrrr}
\toprule
Framework & Configuration & Interval & $u$ & $v$ & SSH & $T$ & $S$ & $\rho$ (EOS)\\
\midrule
Implicit & Biharmonic (physics) & -- & $+10.94$ & $+8.39$ & $+7.60$ & $+6.84$ & $+15.70$ & $+7.37$\\
Implicit & Backscatter (physics) & -- & $+11.44$ & $+8.82$ & $+7.27$ & $+6.76$ & $+16.53$ & $+7.33$\\
Implicit & NeuralPOM (Undamped) & 1 day & $-7.03$ & $-5.27$ & $-6.72$ & $-9.90$ & $-14.72$ & $-7.75$\\
Implicit & NeuralPOM (DND) & 1 day & $-6.45$ & $-4.97$ & $-6.21$ & $-9.59$ & $-16.56$ & $-7.87$\\
Implicit & NeuralPOM (Undamped) & 1 h & $-2.01$ & $-0.53$ & $-3.53$ & $+3.16$ & $+58.66$ & $+3.63$\\
Implicit & NeuralPOM (DND) & 1 h & $-9.47$ & $-7.32$ & $-12.12$ & $-9.04$ & $-0.93$ & $-7.11$\\
\midrule
Explicit & Biharmonic (physics) & -- & $+7.54$ & $+6.29$ & $+5.05$ & $+3.67$ & $+4.60$ & $+3.70$\\
Explicit & Backscatter (physics) & -- & $+8.62$ & $+7.20$ & $+5.69$ & $+4.17$ & $+5.19$ & $+4.20$\\
Explicit & NeuralPOM (Undamped) & 1 day & $-9.62$ & $-4.92$ & $-9.03$ & $-11.83$ & $-16.17$ & $-9.85$\\
Explicit & NeuralPOM (DND) & 1 day & $-9.46$ & $-5.13$ & $-8.96$ & $-11.98$ & $-17.95$ & $-10.56$\\
Explicit & NeuralPOM (Undamped) & 1 h & $-5.69$ & $-3.33$ & $-7.89$ & $+5.38$ & $+90.57$ & $+3.46$\\
Explicit & NeuralPOM (DND) & 1 h & $-10.35$ & $-8.64$ & $-13.49$ & $-8.15$ & $+7.98$ & $-5.63$\\
\bottomrule
\end{tabular}
\end{table}

Comparing low-frequency daily and high-frequency hourly coupling illustrates the trade-off between frequent circulation guidance and baroclinic sensitivity to tracer updates. When deployed with DND, high-frequency hourly coupling (\(\Delta=1\,\mathrm h\)) produces the largest circulation and sea-level error reductions in both frameworks, lowering 100-day RMSE by \(7\%\text{--}10\%\) in horizontal velocity and \(12\%\text{--}13\%\) in SSH alongside \(8\%\text{--}9\%\) reductions in temperature error and \(6\%\text{--}7\%\) reductions in density error. Without directional damping (Undamped), however, high-frequency tracer updates degrade salinity by \(59\%\text{--}91\%\) and increase temperature and density errors by \(3\%\text{--}5\%\) relative to Smagorinsky; as shown in Section~\ref{sec:analysis}, DND prevents this degradation by attenuating the density-modifying component of the tracer increment. On the uncorrected coarse grid, replacing harmonic Smagorinsky diffusion with scale-selective Biharmonic viscosity or kinetic-energy Backscatter increases 100-day RMSE across all variables by \(4\%\text{--}17\%\) in both frameworks (Table~\ref{tab:skill}).

\subsection{Comparison with direct neural forecasters}
\label{sec:results-ai}

We next compare low-frequency daily NeuralPOM (DND) against the five direct neural forecasters evaluated at both their Pretrain (Pre) and Posttrain (Post) stages on the semi-implicit reference (Table~\ref{tab:ai-main}; early- and late-window splits are reported in Appendix~\ref{app:ai-comparison}). Across the 100-day evaluation window, daily NeuralPOM achieves the lowest RMSE in every prognostic field, improving upon the strongest direct neural baseline in each variable by \(5\%\text{--}12\%\) in horizontal velocity and SSH, \(28\%\) in temperature, and \(74\%\) in salinity.

\begin{table}[tbp]
\centering
\caption{\textbf{Comparison of hybrid NeuralPOM and direct neural forecasters.} Pooled 100-day RMSE on the semi-implicit reference across the 16 test origins. Shading and boldface mark the best value in each column and underlining marks the second best, ranked on unrounded scores. Pre and Post denote the frozen Pretrain and Posttrain checkpoints of each direct neural architecture; salinity is in model units.}
\label{tab:ai-main}
\small
\renewcommand{\arraystretch}{1.12}
\setlength{\tabcolsep}{4pt}
\makeatletter\let\AIprimitiveinput\@@input\makeatother
\begin{tabular}{lrrrrr}
\toprule
Configuration & $u$ $\downarrow$ & $v$ $\downarrow$ & SSH $\downarrow$ & $T$ $\downarrow$ & $S$ $\downarrow$\\
 & $\mathrm{m\,s^{-1}}$ & $\mathrm{m\,s^{-1}}$ & $\mathrm m$ & $^\circ\mathrm C$ & \\
\midrule
\AIprimitiveinput data/ai_baselines/table_1_100_rows.tex
\bottomrule
\end{tabular}
\end{table}

For direct neural models trained only with single-step pretraining (Pre), autoregressive errors compound over the 100-day rollout, yielding substantially larger errors than the uncorrected coarse-grid Smagorinsky baseline across all variables. Incorporating 10-day closed-loop rollout training (Post) markedly improves the long-range stability of all five direct neural architectures and brings their 100-day velocity and SSH errors close to the coarse-grid physical baseline. Even after closed-loop training, however, pure data-driven rollouts without an underlying dynamical framework and physical conservation constraints struggle to maintain three-dimensional thermohaline structure over multi-month horizons, incurring larger temperature errors and three- to eightfold higher salinity RMSE than the uncorrected physical solver.

\subsection{Eddy variability, phase alignment and spatial error structure}
\label{sec:results-structure}
\label{sec:results-spectra}

\begin{figure}[tbp]
\centering
\includegraphics[width=\linewidth]{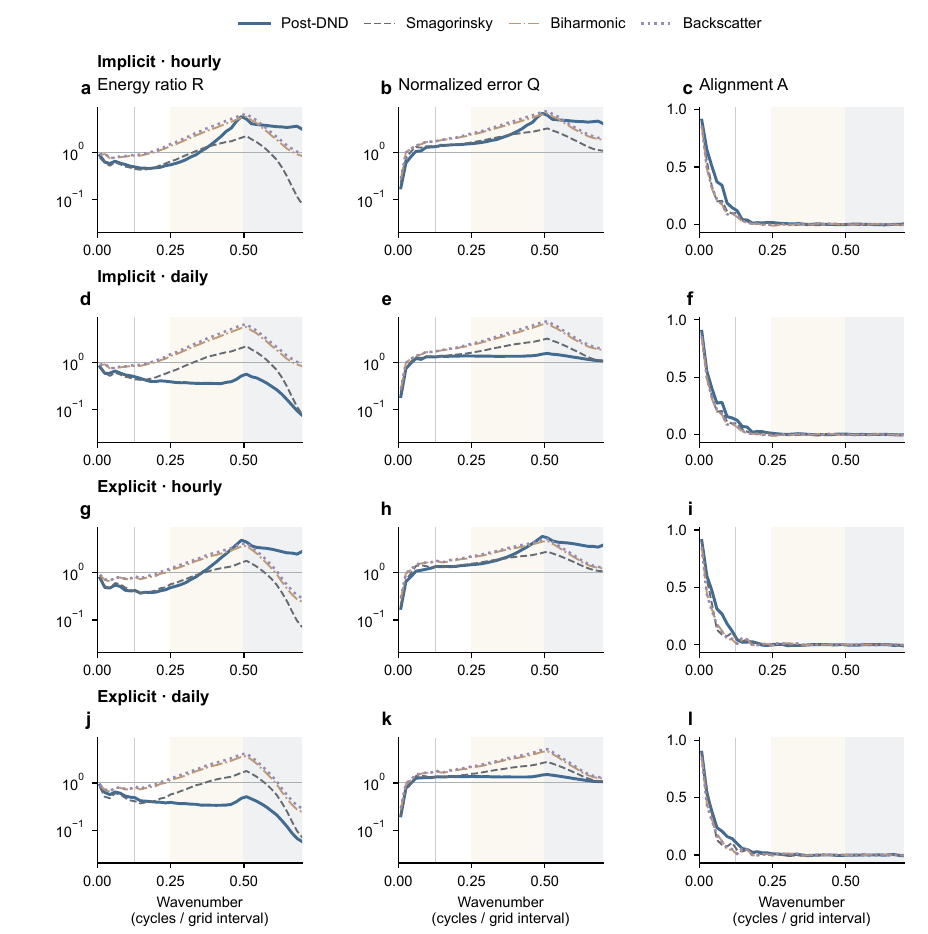}
\caption{\textbf{Wavenumber velocity spectra of the late forecast.} Rows correspond to the four Posttrain hybrid configurations deployed with DND over days 51--100. Columns display the predicted-to-reference kinetic-energy ratio \(R(k)\), the error-to-reference energy ratio \(Q(k)\) and the phase alignment \(A(k)\) as functions of radial wavenumber \(k\) (cycles per grid interval), weighted by layer thickness. Shaded bands denote the near-grid scale (\(0.25<k\le 0.5\text{ cpg}\)) and corner wavenumbers (\(k>0.5\text{ cpg}\)).}
\label{fig:velocity-spectra}
\end{figure}

In turbulent ocean flows, grid-point RMSE mixes eddy amplitude errors with spatial phase errors. To separate these two effects across spatial scales, we evaluate horizontal velocity wavenumber spectra over days 51--100 using three dimensionless metrics (Table~\ref{tab:spectra}, Fig.~\ref{fig:velocity-spectra}, and Appendix~\ref{app:evaluation-spectra}): the energy ratio \(R\) of predicted to reference kinetic energy (\(R=1\) matches the reference energy level; \(R_{\mathrm{near\text{-}grid}}\) denotes its value at near-grid scales), the phase alignment \(A\in[-1,1]\) measuring spatial coherence between predicted and reference eddies (\(A=1\) indicates perfect phase alignment), and the normalized error energy \(Q\) (where lower \(Q\) indicates smaller spectral error). These three quantities satisfy
\begin{equation}
 Q=R+1-2A\sqrt R ,
 \label{eq:theory-spectral}
\end{equation}
which decomposes any change in spectral error relative to the Smagorinsky baseline (\(\Delta Q\)) into an amplitude term \(\Delta Q_{\mathrm{amp}}=\Delta R\) and a phase-alignment term \(\Delta Q_{\mathrm{align}}=-2\Delta(A\sqrt R)\). At small scales where phase alignment is low (\(A\approx 0\), so \(Q\approx 1+R\)), Eq.~\eqref{eq:theory-spectral} also captures the double-penalty effect \cite{Leith1974MonteCarlo,Kochkov2024NeuralGCM}: energizing unaligned eddies increases \(Q\), whereas smoothing out small-scale variability (\(R\to 0\)) lowers \(Q\) at the expense of physical structure.

\begin{table}[tbp]
\centering
\caption{\textbf{Layer-weighted velocity spectral statistics over forecast days 51--100.} Full-scale energy ratio \(R\), phase alignment \(A\), normalized error energy \(Q=R+1-2A\sqrt R\), additive decomposition of \(\Delta Q = Q - Q_{\mathrm{Smag}}\) into amplitude (\(\Delta R\)) and alignment (\(-2\Delta(A\sqrt R)\)) contributions relative to Smagorinsky of the same dynamical framework, and near-grid energy ratio \(R_{\mathrm{near\text{-}grid}}\) (\(0.25<k\le 0.5\text{ cpg}\)). Hybrid rows show the final Posttrain models deployed with DND; direct AI models are evaluated on the semi-implicit reference.}
\label{tab:spectra}
\small
\setlength{\tabcolsep}{4pt}
\begin{tabular}{llrrrrrr}
\toprule
Model & Interval & $R$ & $A$ & $Q$ $\downarrow$ & $\Delta Q_{\mathrm{amp}}$ & $\Delta Q_{\mathrm{align}}$ & $R_{\mathrm{near\text{-}grid}}$\\
\midrule
Smagorinsky (Implicit) & -- & 0.559 & 0.253 & 1.181 & -- & -- & 0.998\\
Biharmonic (Implicit) & -- & 0.819 & 0.229 & 1.404 & $+0.260$ & $-0.037$ & 2.231\\
Backscatter (Implicit) & -- & 0.840 & 0.237 & 1.405 & $+0.281$ & $-0.057$ & 2.512\\
NeuralPOM Implicit (DND) & 1 day & 0.595 & 0.315 & 1.109 & $+0.036$ & $-0.108$ & 0.378\\
NeuralPOM Implicit (DND) & 1 h & 0.604 & 0.385 & 1.006 & $+0.045$ & $-0.220$ & 1.134\\
\midrule
Smagorinsky (Explicit) & -- & 0.528 & 0.222 & 1.205 & -- & -- & 0.836\\
Biharmonic (Explicit) & -- & 0.753 & 0.205 & 1.397 & $+0.226$ & $-0.033$ & 1.700\\
Backscatter (Explicit) & -- & 0.776 & 0.197 & 1.428 & $+0.248$ & $-0.025$ & 1.884\\
NeuralPOM Explicit (DND) & 1 day & 0.581 & 0.287 & 1.143 & $+0.053$ & $-0.115$ & 0.365\\
NeuralPOM Explicit (DND) & 1 h & 0.504 & 0.340 & 1.022 & $-0.024$ & $-0.159$ & 1.033\\
\midrule
Triton Post (direct AI) & 1 day & 0.562 & 0.241 & 1.201 & $+0.003$ & $+0.017$ & 0.072\\
SimVP Post (direct AI) & 1 day & 0.789 & 0.329 & 1.204 & $+0.230$ & $-0.207$ & 0.090\\
ResUNet Post (direct AI) & 1 day & 0.753 & 0.306 & 1.222 & $+0.194$ & $-0.153$ & 0.121\\
ConvLSTM Post (direct AI) & 1 day & 0.600 & 0.160 & 1.352 & $+0.041$ & $+0.129$ & 0.191\\
PastNet Post (direct AI) & 1 day & 0.741 & 0.185 & 1.421 & $+0.181$ & $+0.059$ & 0.294\\
\bottomrule
\end{tabular}
\end{table}

Because the coarse grid (\(\sim 28\,\mathrm{km}\)) physically resolves roughly half of the fine-grid kinetic energy in the reference Smagorinsky solver, an effective hybrid model should maintain this resolved energy level (\(\Delta Q_{\mathrm{amp}}\approx 0\)) while improving the spatial phase alignment \(A\) of mesoscale eddies (\(\Delta Q_{\mathrm{align}}<0\)). As shown in Table~\ref{tab:spectra} and Fig.~\ref{fig:velocity-spectra}, all four NeuralPOM (DND) configurations follow this pattern: they keep bulk kinetic energy \(R\) close to the Smagorinsky baseline while consistently increasing full-spectrum phase alignment \(A\), so that the reduction in spectral error \(Q\) comes almost entirely from the phase-alignment term \(\Delta Q_{\mathrm{align}}\). In addition, high-frequency hourly coupling (\(\Delta=1\,\mathrm h\)) preserves realistic near-grid eddy variability (\(R_{\mathrm{near\text{-}grid}}\approx 1.0\text{--}1.1\)), whereas low-frequency daily coupling (\(\Delta=1\,\mathrm{day}\)) partially smooths near-grid scales.

By contrast, the purely physical and purely data-driven baselines exhibit opposite limitations (Table~\ref{tab:spectra}). On the uncorrected coarse grid, replacing Smagorinsky diffusion with Biharmonic viscosity or kinetic-energy Backscatter injects substantial energy into small scales (raising both \(R\) and \(R_{\mathrm{near\text{-}grid}}\)), but without data-driven guidance purely physical closures cannot align the spatial phase of these eddies; the resulting amplitude penalty (\(\Delta Q_{\mathrm{amp}}>0\)) increases total spectral error \(Q\) relative to Smagorinsky. Conversely, direct neural forecasters trained with closed-loop rollouts (Post) recover large-scale kinetic energy and, in architectures such as SimVP and ResUNet, achieve reasonable large-scale phase alignment, but they strongly damp near-grid variability (\(R_{\mathrm{near\text{-}grid}}<0.3\)) where phase predictability is low, leaving their full-spectrum error \(Q\) higher than both Smagorinsky and NeuralPOM. Complementing the velocity spectra, a spatial decomposition of three-dimensional tracer error (Eq.~\eqref{eq:method-tracer-decomposition}; Fig.~\ref{fig:error-growth} and Appendix~\ref{app:evaluation-tracers}) confirms that horizontal eddy and frontal structure dominates 100-day temperature and salinity errors. Overall, these diagnostics show that NeuralPOM improves long-range skill mainly by better aligning resolved mesoscale eddies and fronts while relying on the physical solver to sustain small-scale variability.

\subsection{Contributions of staged training and route ablation}
\label{sec:results-stages}

We next examine how the three training stages (\(\mathrm{Pretrain}\to\mathrm{Midtrain}\to\mathrm{Posttrain}\)) and three controlled training routes contribute to 100-day forecast skill at matched update budgets (Table~\ref{tab:stage-transitions} and Fig.~\ref{fig:forecast-skill}a; Appendix~\ref{app:training-routes}). Here \emph{Full} denotes the complete three-stage procedure (\(\mathrm{Pretrain\to Midtrain\to Posttrain}\)); \emph{Skip} omits the intermediate stage and proceeds directly from Pretrain to Posttrain (\(\mathrm{Pretrain\to Posttrain}\)); and \emph{Detached} (\(\mathrm{Pretrain\to Midtrain_{\mathrm{detached}}\to Posttrain}\)) executes the same 4-step Midtrain rollouts as \emph{Full} but detaches state gradients at the entrance of each physical integration interval \(\mathcal P^\Delta\), preventing gradients from backpropagating through the ocean solver.

\begin{table}[tbp]
\centering
\caption{\textbf{Staged training progression and controlled route ablations under DND deployment.} For each configuration deployed with DND (\(c=3/4\)), Mid/Pre and Post/Mid report percentage changes in 100-day RMSE between successive stages (negative values indicate error reductions). Full vs.\ Skip and Full vs.\ Detached report relative 100-day RMSE reductions \(100\%\times(1-E_{\mathrm{Full}}/E_{\mathrm{control}})\) of the full route (\(\mathrm{Pre\to Mid\to Post}\)) over skipping Midtrain (\(\mathrm{Pre\to Post}\)) and detached-gradient Midtrain (\(\mathrm{Pre\to Mid_{detached}\to Post}\)) at matched update budgets (positive values favor Full).}
\label{tab:stage-transitions}
\label{tab:routes}
\small
\renewcommand{\arraystretch}{0.96}
\setlength{\tabcolsep}{4.5pt}
\begin{tabular}{lllrrrrr}
\toprule
Framework & Interval & Comparison & $u$ & $v$ & SSH & $T$ & $S$\\
\midrule
Implicit & 1 day & Mid/Pre ($\Delta\%$) & $-3.47$ & $-2.33$ & $-4.62$ & $-3.28$ & $-4.66$\\
 & & Post/Mid ($\Delta\%$) & $+0.19$ & $+0.15$ & $-0.04$ & $-2.07$ & $-1.49$\\
 & & Full vs.\ Skip (\% gain) & $+0.09$ & $+0.72$ & $+0.39$ & $+0.21$ & $+0.34$\\
 & & Full vs.\ Detached (\% gain) & $-0.38$ & $+0.06$ & $-0.20$ & $-0.13$ & $-0.10$\\
\midrule
Explicit & 1 day & Mid/Pre ($\Delta\%$) & $-0.56$ & $-2.28$ & $-1.66$ & $+0.63$ & $-3.38$\\
 & & Post/Mid ($\Delta\%$) & $-4.12$ & $-1.90$ & $-5.50$ & $-8.39$ & $-4.46$\\
 & & Full vs.\ Skip (\% gain) & $+0.29$ & $+0.10$ & $+0.33$ & $+0.33$ & $+0.87$\\
 & & Full vs.\ Detached (\% gain) & $-0.74$ & $-0.54$ & $-1.05$ & $-0.85$ & $-0.40$\\
\midrule
Implicit & 1 h & Mid/Pre ($\Delta\%$) & $+8.39$ & $+7.74$ & $+17.02$ & $+18.70$ & $-15.98$\\
 & & Post/Mid ($\Delta\%$) & $-24.55$ & $-22.23$ & $-29.18$ & $-32.63$ & $-23.28$\\
 & & Full vs.\ Skip (\% gain) & $-1.58$ & $-1.59$ & $-3.71$ & $+1.02$ & $+3.58$\\
 & & Full vs.\ Detached (\% gain) & $+0.06$ & $+0.39$ & $-1.82$ & $+0.96$ & $+9.71$\\
\midrule
Explicit & 1 h & Mid/Pre ($\Delta\%$) & $+12.35$ & $+10.60$ & $+18.87$ & $+19.36$ & $-4.35$\\
 & & Post/Mid ($\Delta\%$) & $-24.00$ & $-23.02$ & $-28.21$ & $-29.17$ & $-17.22$\\
 & & Full vs.\ Skip (\% gain) & $+3.23$ & $+2.67$ & $+3.46$ & $+8.42$ & $+41.15$\\
 & & Full vs.\ Detached (\% gain) & $+4.66$ & $+4.88$ & $+5.09$ & $+8.80$ & $+38.75$\\
\bottomrule
\end{tabular}
\end{table}

Across the three training stages, the behavior of the 4-step differentiable Midtrain stage depends on the physical horizon spanned by those four steps (Table~\ref{tab:stage-transitions}). Under low-frequency daily coupling (\(\Delta=1\,\mathrm{day}\)), four steps span a 4-day window, and both Midtrain and 10-day closed-loop Posttrain reduce 100-day RMSE progressively across both frameworks. Under high-frequency hourly coupling (\(\Delta=1\,\mathrm h\)), however, four steps span only \(4\,\mathrm h\): optimizing over this short window already lowers 100-day salinity error, but it also energizes mesoscale eddies over the short horizon and temporarily elevates 100-day velocity, SSH, and temperature RMSE relative to Pretrain. Subsequent 10-day closed-loop Posttrain (spanning 240 hourly cycles per rollout) damps this excess energy growth and stabilizes the long-range trajectory, reducing 100-day RMSE relative to Midtrain by \(17\%\text{--}33\%\) across all five variables.

Comparing the three training routes (\emph{Full}, \emph{Skip}, and \emph{Detached}) isolates the effect of backpropagating gradients through the differentiable physical solver during Midtrain from multi-step rollout exposure alone (Table~\ref{tab:routes}). Under low-frequency daily coupling, 10-day closed-loop Posttrain is already sufficient to stabilize long-range trajectories, and all three routes converge to nearly identical 100-day skill (within \(1\%\) across all variables). Under high-frequency hourly coupling, by contrast, backpropagating gradients across physical steps plays a key role, especially in the explicit mode-split (\emph{Explicit}) framework where 30 barotropic substeps per baroclinic step propagate fast external gravity waves. In the hourly \emph{Explicit} model, \emph{Detached} offers almost no benefit over \emph{Skip}, whereas retaining gradients through the differentiable solver (\emph{Full}) reduces 100-day velocity and SSH RMSE by \(3\%\text{--}5\%\), temperature RMSE by \(8\%\text{--}9\%\), and salinity RMSE by roughly \(40\%\). A smaller gain in temperature and salinity also appears in the hourly \emph{Implicit} model, where implicit free-surface damping already filters fast external gravity waves during forward integration. In summary, closed-loop Posttrain provides long-range stability across all configurations, while multi-step differentiable Midtrain is critical under high-frequency coupling to account for the short-term dynamical response of the primitive-equation solver.

\section{Baroclinic Feedback of Tracer Corrections and Density-Normal Damping}
\label{sec:analysis}
\label{sec:theory}

\subsection{Baroclinic pressure-gradient feedback and density-normal decomposition}
\label{sec:analysis-frequency}
\label{sec:results-frequency}
\label{sec:analysis-response}
\label{sec:theory-subspaces}
\label{sec:results-shrinkage}
\label{sec:theory-shrinkage}

In Table~\ref{tab:skill}, undamped tracer updates (\(c=1\)) improve all five prognostic variables under low-frequency daily coupling (\(\Delta=1\,\mathrm{day}\)), yet degrade temperature and strongly degrade salinity under high-frequency hourly coupling (\(\Delta=1\,\mathrm h\)). This sensitivity arises because potential temperature \(T\) and salinity \(S\) actively force the three-dimensional primitive equations through the nonlinear equation of state \(\rho(T,S,P)\). Even when a neural update reduces single-step tracer error, uncoordinated small-scale residuals between \(\delta T\) and \(\delta S\) generate local density anomalies \(\delta\rho\) that feed back onto the flow through two physical pathways: horizontally, density gradients \(\nabla_h\delta\rho\) introduce hydrostatic baroclinic pressure-gradient forces that drive near-grid ageostrophic currents; vertically, density gradients \(-\partial_\sigma\delta\rho\) perturb the buoyancy stratification \(N^2\) and modulate the vertical diffusivity and viscosity \((K_H,K_M)\) in the Mellor--Yamada turbulence closure. We hypothesize that under low-frequency daily coupling, the 24-hour physical interval gives the solver sufficient time to disperse and dissipate these transient perturbations between updates, whereas under high-frequency hourly coupling new density anomalies are injected every hour before the flow can adjust, causing near-grid kinetic energy and vertical profile drift to accumulate.

Because both feedback pathways are driven by density perturbations \(\delta\rho\), we can suppress them without sacrificing tracer conservation by separating each layerwise tracer increment according to whether it modifies seawater density (Appendix~\ref{app:theory}). Using the local EOS density gradient on each active \(\sigma\)-layer, volume-weighted orthogonal projection splits any layer-mean-preserving standardized tracer increment \(z=(\delta T'/s_T,\delta S'/s_S)\) into two orthogonal components: a density-modifying (diapycnal) component \(Ez\) aligned with the local density normal, which directly alters buoyancy and baroclinic pressure gradients, and a density-compensated (isopycnal) component \((I-E)z\) tangent to local isopycnals, which adjusts water-mass properties along constant-density surfaces (\(\delta\rho\approx 0\) to first order). Scaling the density-modifying component by \(c\in[0,1]\) yields the filtered update (Eq.~\eqref{eq:method-dnd}) and reduces the squared volume-weighted norm of the density-modifying projection by \(1-c^2\):
\begin{equation}
 z_c=(I-E)z+cEz,\qquad \|z_c\|_w^2=\|z\|_w^2-(1-c^2)\|Ez\|_w^2 .
 \label{eq:theory-retention}
\end{equation}
With \(c=3/4\), Density-Normal Damping (DND) reduces the squared norm of the density-modifying component (and thus the linearized baroclinic pressure-gradient perturbation energy) by \(1-(3/4)^2=43.75\%\), while retaining \(100\%\) of the density-compensated water-mass correction and maintaining exact layerwise heat and salt conservation.

\subsection{Single-perturbation propagation over 72 hours}
\label{sec:analysis-intervention}
\label{sec:theory-frequency}

\begin{figure}[tbp]
\centering
\includegraphics[width=\linewidth]{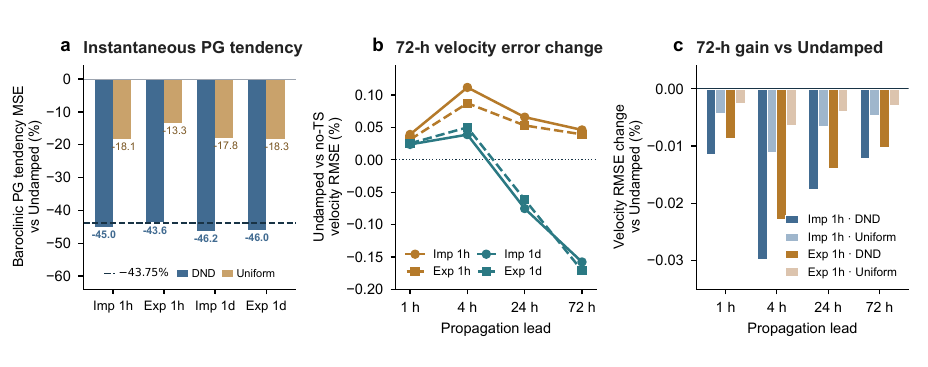}
\caption{\textbf{Single-perturbation physical response and 72-hour propagation across 36 forecast states.} \textbf{a,} Instantaneous percentage change in baroclinic pressure-gradient tendency MSE relative to Undamped at \(t=0\) (dashed line marks the linearized \(-43.75\%\) reduction). \textbf{b,} Percentage change in horizontal velocity RMSE of Undamped relative to omitting the tracer update (no-TS) over 72 hours of free physical integration. \textbf{c,} Percentage change in horizontal velocity RMSE of DND and norm-matched Uniform relative to Undamped in the two hourly models.}
\label{fig:response}
\end{figure}

To test whether DND suppresses baroclinic forcing inside the ocean solver as predicted by Eq.~\eqref{eq:theory-retention}, we design a controlled 72-hour single-perturbation experiment that isolates the physical response to a single neural correction (Fig.~\ref{fig:response} and Appendix~\ref{app:theory-implementation}). Specifically, we first sample 36 pre-correction states from the hybrid NeuralPOM forecast trajectories across all four configurations (three initialization origins at forecast days 1, 10, and 30). At \(t=0\), we invoke the neural corrector once on each state: all four branches receive the identical neural velocity update \((\delta u,\delta v)\), while the neural tracer update \((\delta T,\delta S)\) is applied under four rules---omitting the tracer update (\emph{no-TS}), \emph{Undamped} (\(c=1\)), norm-matched isotropic scaling (\emph{Uniform}), and \emph{DND} (\(c=3/4\)). For \(0<t\le 72\,\mathrm h\), we turn off the neural corrector and integrate all four branches forward using the pure physical solver alone. Omitting subsequent neural updates prevents later corrections from confounding the flow evolution, allowing us to directly trace how the density anomaly \(\delta\rho\) introduced by a single tracer update forces the primitive equations over the following 72 hours.

At \(t=0\), DND reduces the instantaneous baroclinic pressure-gradient tendency MSE relative to \emph{Undamped} by roughly \(45\%\) across all four configurations---closely matching the theoretical \(43.75\%\) reduction and roughly tripling the reduction of norm-matched \emph{Uniform} scaling (Fig.~\ref{fig:response}a)---which in turn cuts tracer-induced horizontal velocity perturbation energy by a similar fraction over the next 72 hours. Tracking horizontal velocity RMSE across the 72-hour window also clarifies the contrast between hourly and daily coupling (Fig.~\ref{fig:response}b,c). In the high-frequency hourly models, a single \emph{Undamped} tracer update excites ageostrophic perturbations that elevate velocity RMSE relative to \emph{no-TS} (Fig.~\ref{fig:response}b), whereas DND suppresses this imbalance and lowers velocity error throughout the 72-hour window (Fig.~\ref{fig:response}c). In the low-frequency daily models, by contrast, the physical flow adjusts over 24--72 hours so that the tracer update reduces velocity error relative to \emph{no-TS} even without strong diapycnal damping.

\subsection{Closed-loop 100-day verification against uniform scaling}
\label{sec:analysis-uniform}

\begin{figure}[tbp]
\centering
\includegraphics[width=\linewidth]{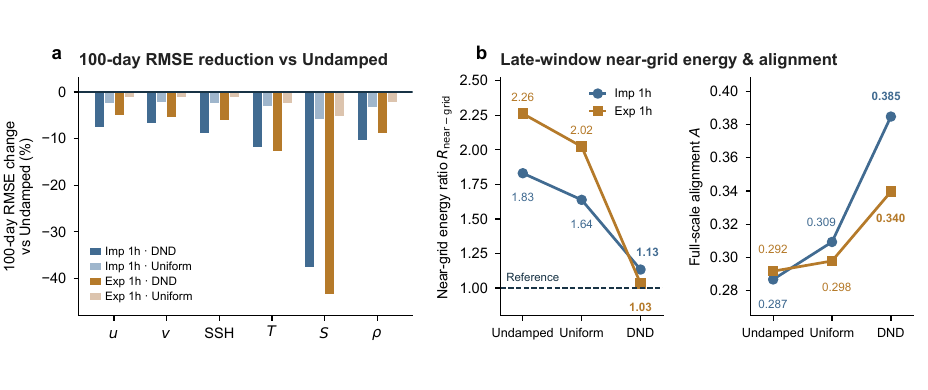}
\caption{\textbf{Closed-loop 100-day verification of DND versus norm-matched Uniform in hourly coupling.} \textbf{a,} Percentage change in 100-day RMSE relative to Undamped on frozen Posttrain networks. \textbf{b,} Late-window (days 51--100) near-grid velocity energy ratio \(R_{\text{near-grid}}\) (left; dashed line marks \(R=1\)) and full-spectrum phase alignment \(A\) (right) across Undamped, Uniform, and DND.}
\label{fig:retention}
\end{figure}

To verify that selectively damping the density-modifying direction---rather than simply shrinking the overall tracer increment---stabilizes long-range forecasts, we compare \emph{Undamped}, norm-matched \emph{Uniform}, and \emph{DND} over 100-day closed-loop integrations from all 16 test origins using identical frozen Posttrain network weights (Table~\ref{tab:retention} and Fig.~\ref{fig:retention}). Under high-frequency hourly coupling, DND lowers 100-day RMSE relative to \emph{Undamped} across all variables, achieving four to eight times the error reductions of norm-matched \emph{Uniform} scaling (Table~\ref{tab:retention} and Fig.~\ref{fig:retention}a). Because all three deployments share the exact same network weights and velocity updates (with no direct neural updates to \(\eta\)), the circulation and sea-level gains under DND result entirely from suppressing spurious baroclinic and vertical-mixing feedbacks during closed-loop cycling.

Late-window diagnostics over days 51--100 confirm that DND stabilizes both physical pathways (Fig.~\ref{fig:retention}b): horizontally, it eliminates the spurious buildup of near-grid kinetic energy and improves full-spectrum phase alignment \(A\); vertically, nearly half of the hourly salinity MSE reduction comes from eliminating vertical profile drift driven by turbulence-closure perturbations. Under low-frequency daily coupling, where the 24-hour physical intervals already allow the flow to adjust between updates, DND leaves 100-day velocity, SSH, and temperature skill virtually unchanged while modestly reducing salinity error (Table~\ref{tab:retention}).

\begin{table}[tbp]
\centering
\caption{\textbf{Closed-loop 100-day comparison of directional damping versus uniform scaling on frozen Posttrain networks.} Percentage change in 100-day RMSE relative to Undamped (\(c=1\)) across the 16 test origins using identical network weights. DND scales only the density-modifying tracer component by \(c=3/4\), whereas Uniform scales the entire tracer increment isotropically to match the mean increment norm of DND.}
\label{tab:retention}
\small
\setlength{\tabcolsep}{4pt}
\begin{tabular}{lllrrrrrr}
\toprule
Framework & Interval & Rule & $u$ & $v$ & SSH & $T$ & $S$ & $\rho$\\
\midrule
Implicit & 1 h & Uniform & $-2.42$ & $-2.24$ & $-2.38$ & $-3.10$ & $-5.87$ & $-3.24$\\
 & & DND & $-7.62$ & $-6.82$ & $-8.90$ & $-11.83$ & $-37.56$ & $-10.37$\\
Explicit & 1 h & Uniform & $-1.10$ & $-1.22$ & $-1.16$ & $-2.48$ & $-5.23$ & $-2.22$\\
 & & DND & $-4.94$ & $-5.50$ & $-6.08$ & $-12.83$ & $-43.34$ & $-8.78$\\
\midrule
Implicit & 1 day & Uniform & $+0.18$ & $+0.04$ & $+0.18$ & $-0.02$ & $-0.18$ & $-0.16$\\
 & & DND & $+0.63$ & $+0.32$ & $+0.55$ & $+0.35$ & $-2.16$ & $-0.13$\\
Explicit & 1 day & Uniform & $+0.21$ & $-0.15$ & $+0.18$ & $-0.07$ & $-0.14$ & $-0.24$\\
 & & DND & $+0.18$ & $-0.22$ & $+0.08$ & $-0.17$ & $-2.13$ & $-0.79$\\
\bottomrule
\end{tabular}
\end{table}

\section{Discussion}
\label{sec:discussion}

NeuralPOM demonstrates that a three-dimensional, terrain-following primitive-equation ocean model and a neural state corrector can be coupled and trained to outperform both coarse-grid subgrid closures and direct neural forecasters over 100-day integrations. On an eddy-permitting coarse grid, conventional subgrid parameterizations cannot recover the spatial phase of unresolved mesoscale eddies from resolved strain rates alone, whereas pure data-driven forecasters damp small-scale variability and struggle to maintain long-range thermohaline conservation. By retaining explicit mode-split and semi-implicit primitive-equation frameworks to enforce volume continuity, dynamical balance, and vertical turbulent mixing while applying conservation-constrained velocity and tracer updates at regular intervals, NeuralPOM lowers forecast errors across all prognostic fields while preserving realistic mesoscale kinetic energy. Achieving this long-range stability requires pairing the constrained inference interface with the three-stage training procedure: closed-loop rollout training (\emph{Posttrain}) suppresses multi-day trajectory drift across all configurations, while multi-step differentiable training (\emph{Midtrain}) teaches the corrector to account for the short-term dynamical response of the ocean solver under high-frequency coupling.

The diagnostic perturbation and closed-loop experiments further clarify how thermohaline state updates interact with primitive-equation dynamics under high-frequency coupling. Because potential temperature and salinity force horizontal momentum and vertical turbulence through the nonlinear equation of state \(\rho(T,S,P)\), uncoordinated residuals in frequent temperature and salinity updates project onto density anomalies, exciting near-grid ageostrophic currents and perturbing vertical mixing before the flow can adjust. Decomposing layerwise tracer increments relative to the local EOS density gradient separates density-compensated (isopycnal) water-mass corrections from density-modifying (diapycnal) perturbations. By selectively attenuating the density-modifying component while preserving the density-compensated component and layerwise conservation, Density-Normal Damping (DND) stabilizes high-frequency coupled forecasts far more effectively than uniform scaling, playing a role analogous to incremental analysis updates \cite{Bloom1996IAU} and balance constraints in operational ocean data assimilation.

The present forecast experiments are conducted in an idealized double-gyre basin with zero surface buoyancy fluxes and complete fine-grid reference states. Extending this framework with probabilistic or spectral training objectives to jointly recover unresolved eddy amplitude and phase \cite{Leith1974MonteCarlo,Kochkov2024NeuralGCM}, and applying the differentiable POM frameworks and constrained coupling interface to realistic regional and global ocean forecasting systems trained against ocean reanalyses or observations, are natural directions for future work.

\section*{Acknowledgements}
This work was supported by the Fundamental and Interdisciplinary Disciplines Breakthrough Plan of the Ministry of Education of China (JYB2025XDXIM801) and the National Natural Science Foundation of China (42125503 and 42430602).

\section*{Data and code availability}
The NeuralPOM source code is available at \url{https://github.com/ChiyodaMomo01/NeuralPOM}. The dataset is available at \url{https://www.modelscope.cn/datasets/chiyodamomo01/NeuralPOM}.

\phantomsection

\clearpage
\appendix\addtocontents{toc}{\protect\setcounter{tocdepth}{1}}\setcounter{section}{0}\renewcommand{\thesection}{\Alph{section}}\renewcommand{\thesubsection}{\thesection.\arabic{subsection}}\renewcommand{\theequation}{\thesection.\arabic{equation}}
\section{Governing equations and numerical discretization}
\label{app:physics}

This appendix specifies the governing equations and discrete operators of NeuralPOM. Both dynamical frameworks solve the hydrostatic, Boussinesq primitive equations in terrain-following coordinates following the Princeton Ocean Model (POM) \cite{BlumbergMellor1987POM} and differ in their free-surface integration: the Explicit framework resolves the external barotropic mode through split-explicit subcycling, whereas the Implicit framework solves a semi-implicit elliptic free-surface system at the internal baroclinic time step. Validation against the double-precision Fortran reference code is presented in Appendix~\ref{app:validation}.

\subsection{Coordinates, variables, and time levels}
\label{app:physics-coordinates}

Horizontal coordinates are \(x,y\), the vertical coordinate \(z\) increases upward, and \(t\) denotes time. With resting bathymetric depth \(h(x,y)>0\) and free-surface elevation \(\eta\), the total water-column thickness is \(H=h+\eta\). Horizontal velocity components are \(u,v\), physical vertical velocity is \(w_z\), and \(g,\rho_0,f\) denote gravitational acceleration, reference seawater density, and the Coriolis parameter. Potential temperature \(T\) and salinity \(S\) are stored relative to reference offsets \((T_{\mathrm{bias}},S_{\mathrm{bias}})\), and density \(\rho\) is represented by the normalized anomaly \(r=(\rho-\rho_{\mathrm{off}})/\rho_0\).

The terrain-following vertical coordinate \(\sigma\in[-1,0]\) and transformed vertical velocity \(\Omega\) are defined by
\begin{equation}
 \sigma=\frac{z-\eta}{H},\qquad z=\eta+\sigma H,\qquad
 \Omega=H\frac{D\sigma}{Dt}=w_z-\partial_t\eta-u\partial_x\eta-v\partial_y\eta-\sigma(\partial_tH+u\partial_xH+v\partial_yH).
 \label{eq:app-omega}
\end{equation}
For any scalar field \(\phi\), horizontal and vertical derivatives transform according to
\begin{equation}
 \left.\partial_x\phi\right|_z=\left.\partial_x\phi\right|_\sigma-\frac{\partial_x\eta+\sigma\partial_xH}{H}\partial_\sigma\phi,\qquad
 \partial_z\phi=\frac{1}{H}\partial_\sigma\phi,
 \label{eq:app-chain-rule}
\end{equation}
and analogously in \(y\). The water column is discretized into \(N_z\) active layers \(k=0,\ldots,N_z-1\) bounded by interfaces \(\sigma_0=0>\sigma_1>\dots>\sigma_{N_z}=-1\), with layer thicknesses \(\Delta\sigma_k=\sigma_k-\sigma_{k+1}\) (\(\sum_{k=0}^{N_z-1}\Delta\sigma_k=1\)), layer centers \(\bar\sigma_k=\tfrac12(\sigma_k+\sigma_{k+1})\), and center-to-center spacings \(\Delta\bar\sigma_k=\bar\sigma_k-\bar\sigma_{k+1}\). An inactive bottom storage level \(k=N_z\) is excluded from vertical integrals.

The internal baroclinic time step is \(\Delta t\) and the external barotropic substep is \(\delta t\). Superscripts \(b,n,f\) denote the backward, current, and forward leapfrog time levels separated by \(2\Delta t\) (internal) or \(2\delta t\) (external), with free-surface pressure weight \(\alpha\) and Robert--Asselin time-filter coefficient \(\mu\). The mode-split formulation tracks external elevation \(\eta_{\mathrm E}\) (\(H_{\mathrm E}=h+\eta_{\mathrm E}\)), internal body elevation \(\eta_{\mathrm B}\) (\(H_{\mathrm B}=h+\eta_{\mathrm B}\)), and the interval-averaged elevation \(\eta_{\mathrm G}\) entering the internal momentum equations.

\subsection{Hydrostatic momentum, continuity, and tracer equations}
\label{app:physics-continuous}
\label{app:physics-tracers}

Let \(p\) denote hydrostatic pressure, \(\eta_a=p_a/(\rho_0g)\) atmospheric pressure head, \(K_M,K_H\) vertical turbulent viscosity and diffusivity, \(\nu\) molecular background diffusivity, and \(\mathcal F_u,\mathcal F_v\) horizontal viscous force divergences. In \(\sigma\)-coordinates, the conservative flux-divergence operator \(\mathscr C(\phi)\) and volume continuity equation take the form
\begin{equation}
 \mathscr C(\phi)=\partial_t(H\phi)+\partial_x(Hu\phi)+\partial_y(Hv\phi)+\partial_\sigma(\Omega\phi),\qquad
 \partial_tH+\partial_x(Hu)+\partial_y(Hv)+\partial_\sigma\Omega=0.
 \label{eq:app-sigma-continuity}
\end{equation}
Integrating continuity over \(\sigma\in[-1,0]\) with an impermeable bottom (\(\Omega|_{\sigma=-1}=0\)), surface volume flux \(F_v\), and depth-averaged velocities \((U,V)=\int_{-1}^0(u,v)\,d\sigma\) yields \(\partial_t\eta+\nabla_h\cdot(HU,HV)+F_v=0\). Writing \(z_x=\partial_x\eta+\sigma\partial_xH\) and \(z_y=\partial_y\eta+\sigma\partial_yH\), the horizontal momentum and hydrostatic relations are
\begin{align}
 \mathscr C(u)-Hfv
 &=-\frac{H}{\rho_0}\Bigl(\partial_xp-\frac{z_x}{H}\partial_\sigma p\Bigr)
   +\partial_\sigma\!\Bigl[\frac{K_M+\nu}{H}\partial_\sigma u\Bigr]+H\mathcal F_u,
 \label{eq:app-momentum-x}\\
 \mathscr C(v)+Hfu
 &=-\frac{H}{\rho_0}\Bigl(\partial_yp-\frac{z_y}{H}\partial_\sigma p\Bigr)
   +\partial_\sigma\!\Bigl[\frac{K_M+\nu}{H}\partial_\sigma v\Bigr]+H\mathcal F_v,
 \qquad \partial_\sigma p=-gH\rho.
 \label{eq:app-sigma-momentum}
\end{align}
Potential temperature \(T\) and salinity \(S\) evolve under horizontal diffusion along \(\sigma\)-surfaces (acting on backward-level anomalies relative to reference climatology \(T_c,S_c\) with turbulent Prandtl ratio \(A_H/A_M=0.20\)) and vertical turbulent diffusion:
\begin{equation}
 \mathscr C(T)=\nabla_h\cdot\bigl[HA_H\nabla_h(T-T_c)\bigr]+\partial_\sigma\!\Bigl[\frac{K_H+\nu}{H}\partial_\sigma T\Bigr]+H\mathcal Q_T,\qquad
 \mathscr C(S)=\nabla_h\cdot\bigl[HA_H\nabla_h(S-S_c)\bigr]+\partial_\sigma\!\Bigl[\frac{K_H+\nu}{H}\partial_\sigma S\Bigr]+H\mathcal Q_S.
 \label{eq:app-tracers-combined}
\end{equation}
Vertical momentum boundary conditions prescribe surface wind stress \((K_M+\nu)\partial_z(u,v)|_{z=\eta}=\boldsymbol\tau_s/\rho_0\) and quadratic bottom drag \((K_M+\nu)\partial_z(u,v)|_{z=-h}=C_D|\boldsymbol u_b|\boldsymbol u_b\). In-situ density \(\rho\) and empirical sound speed \(c\) are evaluated from the nonlinear POM equation of state \cite{BlumbergMellor1987POM,Mellor1991EOS} as functions of potential temperature \(\vartheta=T+T_{\mathrm{bias}}\) (\({}^\circ\mathrm C\)), salinity \(s=S+S_{\mathrm{bias}}\), and static reference pressure \(P=-\rho_0g\bar\sigma h\times 10^{-5}\) (\(\mathrm{bar}\)):
\begin{equation}
 \rho=\rho(\vartheta,s,P),\qquad r=\frac{\rho(\vartheta,s,P)-\rho_{\mathrm{off}}}{\rho_0}.
 \label{eq:app-eos}
\end{equation}

\subsection{Arakawa-C spatial discretization and Smagorinsky viscosity}
\label{app:physics-cgrid}
\label{app:physics-tracer-discrete}

Horizontal operators are discretized on an orthogonal Arakawa-C grid where scalar cell \((i,j)\) has grid spacings \(\Delta x_{ij},\Delta y_{ij}\), area \(A_{ij}=\Delta x_{ij}\Delta y_{ij}\), and wet mask \(m_{T,ij}\), while face velocities \(u_{ij}\) (between cells \(i-1\) and \(i\)) and \(v_{ij}\) (between cells \(j-1\) and \(j\)) carry control areas \(A^u_{ij},A^v_{ij}\), interpolated depths \(H^u_{ij}=\tfrac12(H_{ij}+H_{i-1,j})\) and \(H^v_{ij}=\tfrac12(H_{ij}+H_{i,j-1})\), transverse face lengths \(L^u_{ij},L^v_{ij}\), and face masks \(m_{U,ij},m_{V,ij}\). For masked face transports \(Q^u=m_UH^uu\) and \(Q^v=m_VH^vv\), the discrete horizontal flux divergence is
\begin{equation}
 \bigl(D_h(Q^u,Q^v)\bigr)_{ij}
 =\frac{L^u_{i+1,j}Q^u_{i+1,j}-L^u_{ij}Q^u_{ij}+L^v_{i,j+1}Q^v_{i,j+1}-L^v_{ij}Q^v_{ij}}{A_{ij}}.
 \label{eq:app-flux-divergence}
\end{equation}
Second-order centered differences discretize advective momentum and tracer fluxes at time level \(n\). Horizontal viscous and diffusive fluxes act on backward-level (\(b\)) velocity strain and climatological tracer anomalies \(\phi'^b=\phi^b-\phi_c\) using resting face depths \(h^u,h^v\), advancing tracers to the intermediate state \(\phi^*_{ij,k}=(A_{ij}H^b_{ij}\phi^b_{ij,k}-2\Delta t\,\mathcal L_{\phi,ij,k})/(A_{ij}H^f_{ij})\) prior to implicit vertical diffusion. Horizontal eddy viscosity follows the deformation-dependent Smagorinsky closure with cell-centered normal strain rates \(D_xu,D_yv\) and interpolated shear \(s_{xy}\):
\begin{equation}
 A_M=C_s\,\Delta x\,\Delta y\,\sqrt{(D_xu)^2+(D_yv)^2+\tfrac12s_{xy}^2},\qquad C_s=0.05.
 \label{eq:app-smagorinsky}
\end{equation}

\subsection{Discrete hydrostatic baroclinic pressure gradient}
\label{app:physics-pressure-gradient}

Let \(r'=r-r_{\mathrm{mean}}\) denote the normalized density anomaly relative to the horizontally uniform reference profile \(r_{\mathrm{mean}}\). For the two scalar columns adjacent to a \(u\)-face, define \(\Delta r'_k=r'_{i,k}-r'_{i-1,k}\) and \(\Sigma r'_k=r'_{i,k}+r'_{i-1,k}\). Vertical integration of the \(\sigma\)-transformed hydrostatic relation yields the discrete baroclinic pressure-gradient recurrence
\begin{align}
 B^u_0
 &=\frac{g}{2}(-\bar\sigma_0)(H_i+H_{i-1})\Delta r'_0,
 \label{eq:app-pg-top}\\
 B^u_k
 &=B^u_{k-1}
 +\frac{g}{4}(\bar\sigma_{k-1}-\bar\sigma_k)(H_i+H_{i-1})(\Delta r'_k+\Delta r'_{k-1})
 +\frac{g}{4}(\bar\sigma_{k-1}+\bar\sigma_k)(H_i-H_{i-1})(\Sigma r'_k-\Sigma r'_{k-1}).
 \label{eq:app-pg-recurrence}
\end{align}
The corresponding horizontal baroclinic force is \(\mathcal P^u_k=H^uL^uB^u_km_U\) (and analogously \(\mathcal P^v_k\) in \(y\)). Equations~\eqref{eq:app-pg-top}--\eqref{eq:app-pg-recurrence} enter both the internal momentum solver and the baroclinic pressure-gradient term of the training loss (Appendix~\ref{app:training-loss}).

\subsection{Vertical turbulence closure and implicit column diffusion}
\label{app:physics-turbulence}
\label{app:physics-vertical}

Vertical turbulent mixing is parameterized by the Mellor--Yamada level-2.5 closure \cite{MellorYamada1982} with prognostic equations for twice the turbulent kinetic energy \(q^2\) and the master length-scale product \(q^2l\):
\begin{align}
 \mathscr C(q^2)
 &=\nabla_h\cdot(HA_M\nabla_hq^2)+\partial_\sigma\!\Bigl[\frac{K_Q+\nu}{H}\partial_\sigma q^2\Bigr]+2H\Bigl(P_s+B-\frac{q^3}{B_1l}\Bigr),\\
 \mathscr C(q^2l)
 &=\nabla_h\cdot\bigl[HA_M\nabla_h(q^2l)\bigr]+\partial_\sigma\!\Bigl[\frac{K_Q+\nu}{H}\partial_\sigma(q^2l)\Bigr]+H\Bigl(E_1l(P_s+B)-\frac{q^3}{B_1}W_l\Bigr),
 \label{eq:app-tke-continuous}
\end{align}
where \(P_s=K_M[(\partial_zu)^2+(\partial_zv)^2]\) is vertical shear production, \(B=K_Hb\) is buoyancy production, \(W_l=1+E_2[l\kappa^{-1}((z+h)^{-1}+(\eta-z)^{-1})]^2\) is the wall-proximity function, and \((B_1,E_1,E_2,\kappa)=(16.6,1.8,1.33,0.40)\). At interface \(k\), with compressibility-corrected sound speed \(c_k\) from the POM equation of state, the vertical buoyancy gradient \(b_k\), stability parameter \(G_H\), and dissipation coefficient \(d_k\) are given by
\begin{equation}
 b_k=\frac{g(r_{k-1}-r_k)}{\Delta\bar\sigma_{k-1}H}+\frac{2g^2}{c_{k-1}^2+c_k^2},\qquad
 G_H=\min\!\Bigl(\frac{l_k^2b_k}{|q_b^2|},\,0.028\Bigr),\qquad
 d_k=\frac{\sqrt{|q_b^2|}}{B_1l_k+\epsilon_l}.
 \label{eq:app-tke-buoyancy}
\end{equation}
With closure constants \((A_1,A_2,B_2,C_1)=(0.92,0.74,10.1,0.08)\), the stability functions \(S_H(G_H),S_M(G_H)\) determine the updated vertical diffusivities \(K_M,K_H,K_Q\), which are averaged equally with their values from the preceding step. Boundary conditions for \(q^2\) and \(q^2l\) are set by surface and bottom kinematic stress magnitudes \(a_s,a_b\):
\begin{align}
 l_0&=2\times 10^5a_s/g,\qquad Q_s=(15.8\times 100)^{2/3}a_s,\qquad Q_{\mathrm{bot}}=B_1^{2/3}a_b,
 \label{eq:app-tke-wall-energy}\\
 L_1&=-\kappa\sigma_1H_fq_1^2,\qquad L_{N_z-1}=\kappa(1+\sigma_{N_z-1})H_fq_{N_z-1}^2.
 \label{eq:app-tke-wall-length}
\end{align}
Vertical diffusive updates for \((u,v,T,S,q^2,q^2l)\) are solved implicitly over \(2\Delta t\) via tridiagonal column inversions across the active layers, with quadratic bottom drag \(\gamma_b=C_D|\boldsymbol u_b^b|\) entering the bottom momentum diagonal and logarithmic bottom drag coefficient
\begin{equation}
 C_D=\operatorname{clip}\!\left[\left\{\frac{\kappa}{\ln[(z_{\mathrm{sh}}+(1+\bar\sigma_{\mathrm{bottom}})H_{\mathrm E})/z_0]}\right\}^2,\ C_{D,\min},\,C_{D,\max}\right],
 \label{eq:app-coastal-drag}
\end{equation}
where \(C_{D,\min}=0.0025\) and \(C_{D,\max}=1\).

\subsection{Explicit mode-split integration}
\label{app:physics-explicit}

With mode-split step ratio \(N_e=\Delta t/\delta t\), the Explicit framework advances the external barotropic mode over \(m=0,\dots,N_e-1\) substeps driven by depth-averaged baroclinic forcing held fixed across the internal step:
\begin{align}
 \eta^{m+1}&=\eta^{m-1}-2\delta t\bigl[D_h(H^{u,m}U^m,H^{v,m}V^m)+F_v^m\bigr],
 \label{eq:app-explicit-elevation}\\
 H^{u,m+1}U^{m+1}&=H^{u,m-1}U^{m-1}-\frac{2\delta t}{A^u}\Bigl[\overline{\mathcal A}^u-A^uC^u+gH^uL^u\delta_x\chi+\overline{\mathcal P}^u+A^u(a_s^u-a_b^u)\Bigr],
 \label{eq:app-explicit-momentum}
\end{align}
where \(\chi=(1-2\alpha)\eta^m+\alpha(\eta^{m-1}+\eta^{m+1})+\eta_a\), \(C^u\) is the depth-weighted Coriolis term, and \(V^{m+1}\) is updated analogously. At each step, leapfrog levels are damped by the Robert--Asselin filter \(\widetilde x^n=x^n+\tfrac{\mu}{2}(x^b-2x^n+x^f)\) (with the vertical mean removed from \(\Delta^2u_k,\Delta^2v_k\) so that three-dimensional filtering preserves the barotropic transport). Across the \(N_e\) substeps, the solver accumulates the forward body elevation \(\eta_{\mathrm B}^f=\tfrac{\mu}{4}\eta_f^{(N_e-2)}+\tfrac12(1-\tfrac{\mu}{2})\eta_f^{(N_e-1)}+\tfrac12\eta_f^{(N_e)}\), the interval-averaged elevation \(\eta_{\mathrm G}^f\), and the mean external transports \(\mathcal Q_+^{u,v}=N_e^{-1}\sum_{m=0}^{N_e-1}H^{u,v,(m)}(U,V)^{(m)}\).

\subsection{Semi-implicit free-surface solver and implicit adjoint}
\label{app:physics-implicit}

The Implicit framework advances the external mode directly over \(\Delta t\). Forming predictor transports \(Q_*^u=H^{u,n}U_*=H^{u,b}U^b-(2\Delta t/A^u)\mathcal N_*^u\) (and \(Q_*^v\) analogously) with explicit pressure head \(\chi_*=(1-2\alpha)\eta^n+\alpha\eta^b+\eta_a\) and right-hand side \(b_\eta=\eta^b-2\Delta t[D_h(Q_*^u,Q_*^v)+F_v^f]\), the forward free-surface elevation \(\eta^f\) satisfies the elliptic system
\begin{equation}
 L_H\phi=D_h(m_UH^uG_x\phi,\,m_VH^vG_y\phi),\qquad
 A_\eta=I-4\Delta t^2g\alpha L_{H^n},\qquad
 A_\eta\eta^f=b_\eta.
 \label{eq:app-elliptic}
\end{equation}
The velocity corrector updates the barotropic transport via \(H^{u,f}U^f=H^{u,n}U_*-2\Delta t\,g\alpha H^{u,f}G_x\eta^f\) in the double-gyre forecast experiments and via the conservative form
\begin{equation}
 H^{u,f}U^f=H^{u,n}U_*-2\Delta t\,g\alpha H^{u,n}G_x\eta^f
 \label{eq:app-conservative-corrector}
\end{equation}
in the Fortran validation suite. The linear system \(A_\eta\eta^f=b_\eta\) is solved by preconditioned conjugate gradients (PCG) with a fast discrete sine transform (DST-I) spectral preconditioner \(\mathcal S\) based on the constant-coefficient Helmholtz eigenvalues \(\lambda_{ab}=4\Delta x_0^{-2}\sin^2[\pi a/(2(n_x+1))]+4\Delta y_0^{-2}\sin^2[\pi b/(2(n_y+1))]\) evaluated at the median resting depth \(H_0\),
\begin{equation}
 B^{-1}r=\mathcal S^{-1}\!\left[\frac{\mathcal S(r)}{1+4\Delta t^2g\alpha H_0\lambda_{ab}}\right].
 \label{eq:app-spectral-preconditioner}
\end{equation}

To propagate sensitivities through the elliptic solve without unrolling PCG iterations, we apply implicit differentiation. For any state or parameter \(\zeta\) entering \(A_\eta\) or \(b_\eta\) and scalar objective \(\mathcal L\) with upstream gradient \(g_\eta=\partial\mathcal L/\partial\eta^f\), differentiating \(A_\eta\eta^f=b_\eta\) gives
\begin{equation}
 A_\eta^\top\lambda=g_\eta,\qquad
 \frac{d\mathcal L}{d\zeta}=\lambda^\top\left[\frac{\partial b_\eta}{\partial\zeta}-\frac{\partial A_\eta}{\partial\zeta}\eta^f\right].
 \label{eq:app-implicit-adjoint}
\end{equation}
Because \(A_\eta\) is self-adjoint with respect to the cell-area inner product \(M_A=\operatorname{diag}(A_{ij})\), \(A_\eta^\top=M_AA_\eta M_A^{-1}\) and the adjoint variable is obtained from a single PCG solve:
\begin{equation}
 \lambda=M_AA_\eta^{-1}(M_A^{-1}g_\eta).
 \label{eq:app-area-adjoint}
\end{equation}

\subsection{Projection and three-dimensional mode coupling}
\label{app:physics-coupling}

Prior to internal baroclinic transport, horizontal velocities are constrained to match the two-interval average external transport \(\overline{\mathcal Q}^{u,v}=\tfrac12(\mathcal Q_-^{u,v}+\mathcal Q_+^{u,v})\) while preserving vertical shear, \(u_k\leftarrow u_k-\sum_\ell\Delta\sigma_\ell u_\ell+\overline{\mathcal Q}^u/H_{\mathrm B}^u\) (and analogously for \(v_k\)). The transformed vertical velocity \(\Omega\) is then diagnosed downward from the surface condition \(\Omega_0=\overline F_v\) via discrete continuity:
\begin{equation}
 \Omega_{k+1}=\Omega_k+\Delta\sigma_k\left[D_h(H_{\mathrm B}^uu_k,H_{\mathrm B}^vv_k)+\frac{\eta_{\mathrm B}^f-\eta_{\mathrm B}^b}{2\Delta t}\right].
 \label{eq:app-vertical-continuity}
\end{equation}
Three-dimensional horizontal momentum is subsequently advanced with surface pressure gradient \(\delta_x\chi_3\) (\(\chi_3=\tfrac12(\eta_{\mathrm G}^f+\eta_{\mathrm G}^b)+\eta_a\)) and baroclinic pressure-gradient force \(\mathcal P_k^{u,v}\), followed by the implicit vertical viscosity solve, the vertical-mean-preserving Robert--Asselin filter, and the leapfrog time-level shift.

\section{State Correction Interface, Objective, and Training}
\label{app:training}

\subsection{Dataset construction, splits, and target alignment}
\label{app:training-data}

For each dynamical framework (Explicit and Implicit), the fine-grid reference ensemble comprises eight members integrated for 9,000 model days each on a \(354\times 322\times 16\) grid (\(\sim 7\,\mathrm{km}\) horizontal spacing, \(352\times 320\) wet interior cells, 15 active \(\sigma\)-layers plus one inactive bottom storage level) from distinct divergence-free velocity perturbations, with potential temperature initialized from an exponential vertical stratification profile and salinity from a uniform value of 35 in model units. Volume-weighted conservative restriction onto the \(90\times 82\times 16\) coarse forecast grid (\(\sim 28\,\mathrm{km}\) spacing, \(88\times 80\) wet interior cells) archives complete daily 45-variable state checkpoints alongside 20-minute subdaily targets during the first four hours of each day. Ensemble members 0--6 over days 1000--7000 provide training trajectories, member 7 over days 7000--8000 provides validation data, and out-of-sample test evaluations are initialized from member 7 at 16 evenly spaced forecast origins \(d_r = 8000 + 32r\) (\(r=0,\dots,15\)) and integrated for 100 days. Each forecast initialized at start day \(d\) begins from the fine-grid reference state projected onto the coarse grid at the end of day \(d-1\). Daily reference targets correspond to day-boundary states; hourly reference targets are extracted from the 20-minute reference fields archived during the first four hours of each day, so that each one-hour correction interval spans six 600-s coarse-grid internal steps (three 20-minute reference intervals). An offline cache stores the uncorrected coarse-grid state after the first physical integration interval, providing the exact pre-correction state for single-step pretraining and for the first coupling step of multi-step differentiable training.

\subsection{Neural corrector architecture}
\label{app:training-network}

The network input concatenates the three-dimensional current and backward leapfrog horizontal velocities and tracers \((u,u^b,v,v^b,T,S)\) across all 16 stored vertical levels together with five two-dimensional external-mode fields (free-surface elevation and current and backward depth-averaged velocities), yielding \(6\times16+5=101\) input channels, each standardized by frozen training-set mean and standard deviation statistics. The architecture is a two-dimensional residual U-Net with base width 256 and three \(2\times2\) max-pooling stages (512, 1024, and 2048 channels), paired with a bilinear-upsampling decoder and skip connections. Each residual block consists of two bias-free \(3\times3\) convolutions with group normalization and ReLU activations, supplemented by a \(1\times1\) projection shortcut when channel width changes. The final \(1\times1\) output convolution is initialized to zero so that training begins from an exact zero-increment identity mapping. The network contains \(156\times10^6\) parameters and executes in 32-bit floating point with activation checkpointing during multi-step training.

\subsection{Increment constraints and state update}
\label{app:training-correction}

The 60 output channels (four three-dimensional fields across the 15 active vertical layers) are scaled to physical units via \(\delta f=s_f^{\mathrm{out}}a_{\theta,f}\) for \(f\in\{u,v,T,S\}\). With horizontal cell area \(A_{ij}\), scalar wet mask \(m_{ij}\), resting depth \(h_{ij}\), predicted body elevation \(\eta_{\mathrm B,ij}\), and layer thickness \(\Delta\sigma_k\), the cell-volume weights are \(w_{ijk}=A_{ij}m_{ij}(h_{ij}+\eta_{\mathrm B,ij})\Delta\sigma_k\). At every correction step, tracer increments are projected to zero layerwise volume mean and horizontal velocity increments are projected to zero column mean:
\begin{equation}
 \delta T'_{ijk}=\delta T_{ijk}-\frac{\sum_{ij}w_{ijk}\delta T_{ijk}}{\sum_{ij}w_{ijk}},\qquad
 \delta u'_{ijk}=m_{U,ij}\Bigl[\delta u_{ijk}-\frac{\sum_k\Delta\sigma_k\delta u_{ijk}}{\sum_k\Delta\sigma_k}\Bigr],
 \label{eq:app-correction-projections}
\end{equation}
and analogously for \(S\) and \(v\). When Density-Normal Damping (DND) is applied, the layer-mean-preserving orthogonal decomposition and density-normal attenuation \(z_c=(I-E)z+cEz\) (\(c=3/4\), Appendix~\ref{app:theory}) act on the standardized layer-mean-preserving tracer increment pair \((\delta T'/s_T,\delta S'/s_S)\). Each variable pair is then bounded by a domain-wide scalar limiter \(c_{TS}=z_{\max}/\max(z_{\max},p_{TS})\), where \(p_{TS}\) is the maximum absolute standardized tracer increment across both tracers and all active wet cells, and \(c_{uv}\) analogously for horizontal velocity, with \(z_{\max}=5\). In the state update, the resulting physical increments are added identically to both the current and backward leapfrog time levels, \(f^n\leftarrow f^n+\widehat{\delta f}\) and \(f^b\leftarrow f^b+\widehat{\delta f}\), the inactive bottom level \(k=15\) is left unchanged, and in-situ density \(\rho\) is updated across all wet cells via the nonlinear equation of state.

\subsection{Composite endpoint loss}
\label{app:training-loss}

For variable \(f\) with predicted-geometry volume weights \(w^f\) and error normalization scale \(s_f^{\mathrm{err}}\), define the normalized mean squared error \(N_f=\sum w^f(f-f^\star)^2/\bigl(\sum w^f\,(s^{\mathrm{err}}_f)^2\bigr)\). The momentum term averages current and backward leapfrog velocity errors, \(N_m=\tfrac14(N_u+N_{u^b}+N_v+N_{v^b})\). The baroclinic pressure-gradient consistency term \(N_p\) applies the discrete hydrostatic pressure-gradient recurrence (Eqs.~\eqref{eq:app-pg-top}--\eqref{eq:app-pg-recurrence}) to the equation-of-state density error \(r_e=r-r^\star\), converts the resulting horizontal baroclinic force error \((P_x,P_y)\) into an equivalent single-step leapfrog velocity increment,
\begin{equation}
 k^u_{ijk}=-\frac{4\Delta t\,P_{x,ijk}}{(H_{ij}+H_{i-1,j})A^u_{ij}}\,m_{U,ij},
 \label{eq:app-pressure-response}
\end{equation}
and analogously for \(k^v\), and penalizes \(N_p=\tfrac12[M(k^u)+M(k^v)]/(s_p^{\mathrm{err}})^2\). The horizontal frontal-gradient term \(N_F\) averages the normalized squared differences of temperature and salinity errors between adjacent wet cells in both horizontal grid directions. The composite endpoint loss is
\begin{equation}
 \ell_\chi=0.25N_m+0.25N_T+0.25N_S+0.15N_p+0.10N_F+0.10\,\chi N_\eta,
 \label{eq:app-endpoint-objective}
\end{equation}
where \(\chi=0\) during Pretrain and \(\chi=1\) during Midtrain and Posttrain. Because the state update does not modify free-surface elevation directly, \(N_\eta\) contributes a nonzero parameter gradient only during Midtrain, where earlier velocity and tracer increments influence subsequent sea-surface height through the differentiable primitive-equation solver.

\subsection{Optimization schedule and selected checkpoints}
\label{app:training-optimization}
\label{app:training-stages}
\label{app:training-sensitivity}
\label{app:training-routes}

The three-stage training procedure progresses sequentially through Pretrain, Midtrain, and Posttrain. Pretrain optimizes single-step intervals (\(N=1\)) on cached physical forecasts; Midtrain unrolls \(N=4\) coupled correction-and-integration intervals end-to-end through the differentiable dynamical framework; and Posttrain integrates 10-day closed-loop trajectories, performing one parameter update at the final correction of each day while detaching the computational graph between days. All three stages use eight-GPU data-parallel optimization with AdamW (weight decay \(10^{-4}\)) and global gradient norm clipping at 10. Pretrain samples initialization states every six days across the seven 6000-day training reference members (7,000 samples per epoch) and trains for 40 epochs with a cosine learning-rate schedule decaying from \(5\times10^{-5}\) to \(5\times10^{-6}\) after a three-epoch linear warmup. Midtrain and Posttrain use a fixed learning rate of \(3\times10^{-6}\). Hourly Midtrain samples trajectories every 24 days for 10 epochs; daily Midtrain samples every 72 days for three epochs. Posttrain samples 10-day rollouts every 60 days. Table~\ref{tab:app-training-checkpoints} lists the completed optimizer updates for all four configurations; in the route ablations (Section~\ref{sec:results-stages}), \emph{Skip} omits Midtrain while \emph{Detached} applies a stop-gradient operator at the entrance of each Midtrain physical interval \(\mathcal P^\Delta\) under identical update budgets.

\begin{table}[tb]
\centering\small
\caption{\textbf{Optimizer updates across training stages.} Entries report the approximate number of optimizer updates (rounded to the nearest hundred) for each training stage.}
\label{tab:app-training-checkpoints}
\begin{tabular}{lrrr}
\toprule
Configuration & Pretrain & Midtrain & Posttrain\\
\midrule
Implicit, 1 h & 35,000 & 2,200 & 1,800\\
Implicit, 1 day & 35,000 & 200 & 5,000\\
Explicit, 1 h & 35,000 & 2,200 & 1,800\\
Explicit, 1 day & 35,000 & 200 & 5,000\\
\bottomrule
\end{tabular}
\end{table}

\section{Layer-Constrained Density-Normal Projection}
\label{app:theory}

This appendix derives the layer-mean-preserving orthogonal projection used in Density-Normal Damping (DND, Section~\ref{sec:theory}) and summarizes its numerical implementation.

\subsection{Orthogonal projection onto the density-modifying subspace}
\label{app:theory-projection}
\label{app:theory-ridge}
\label{app:theory-response}
\label{app:theory-risk}
\label{app:theory-frequency}
\label{app:theory-transport}
\label{app:theory-spectral}

Consider a single active \(\sigma\)-layer with wet cells \(i\), normalized cell-volume weights \(w_i>0\) (\(\sum_iw_i=1\)), and the Hilbert space of layer-mean-preserving standardized tracer increments \(\mathcal H_0=\{u=(u_i):\;u_i\in\mathbb R^2,\;\sum_iw_iu_i=0\}\) equipped with the volume-weighted inner product \(\langle u,v\rangle_w=\sum_iw_iu_i^\top v_i\). Let \(r_i=(\rho_i-\rho_{\mathrm{off}})/\rho_0\) denote the normalized in-situ density anomaly at the pre-correction state, \(g_i=(s_T\partial_Tr_i,\,s_S\partial_Sr_i)^\top\) its gradient in standardized tracer coordinates, \(n_i=g_i/\|g_i\|_2\) the local unit density-normal direction, and \(a_i=n_i^\top z_i\) the pointwise density-normal component of a given tracer increment \(z\in\mathcal H_0\).

Because \(n_i\) varies horizontally across the basin due to the nonlinear equation of state, the pointwise density-normal field \(a_in_i\) does not in general have zero layer mean. The subspace of layer-mean-preserving density-modifying (diapycnal) increments is
\begin{equation}
 \mathcal S=\Bigl\{p\in\mathcal H_0:\;p_i=\beta_in_i,\quad \sum_iw_in_i\beta_i=0\Bigr\}.
\end{equation}
Finding the closest element \(p\in\mathcal S\) to \(z\) in the \(\|\cdot\|_w\) norm reduces to minimizing \(\tfrac12\sum_iw_i(\beta_i-a_i)^2\) subject to \(\sum_iw_in_i\beta_i=0\). Introducing a Lagrange multiplier \(\lambda\in\mathbb R^2\), the stationarity conditions give \(\beta_i=a_i-n_i^\top\lambda\) and \(M\lambda=b\), where \(M=\sum_iw_in_in_i^\top\in\mathbb R^{2\times2}\) is the symmetric positive semidefinite Gram matrix of the layer density normals and \(b=\sum_iw_in_ia_i\in\mathbb R^2\). Because any \(t\in\ker M\) satisfies \(0=t^\top Mt=\sum_iw_i(n_i^\top t)^2\) and hence \(n_i^\top t=0\) for all \(i\), we have \(t^\top b=0\), so \(b\in\operatorname{ran}M\). Applying the Moore--Penrose pseudoinverse \(M^+\) \cite{Penrose1955GeneralizedInverse} yields the unique minimum-norm multiplier \(\lambda=M^+b\) and the projected scalar amplitude
\begin{equation}
 \beta_i=a_i-n_i^\top M^+b,\qquad \sum_iw_ip_i=b-MM^+b=0.
 \label{eq:app-theory-amplitude}
\end{equation}
For any test field \(t_i=\gamma_in_i\in\mathcal S\), \(\langle z-p,t\rangle_w=\sum_iw_i(a_i-\beta_i)\gamma_i=\lambda^\top\sum_iw_in_i\gamma_i=0\). Thus \(p=Ez\) defines the exact \(w\)-orthogonal projection of \(\mathcal H_0\) onto \(\mathcal S\), while \((I-E)z\in\mathcal S^\perp\cap\mathcal H_0\) is the layer-mean-preserving density-compensated (isopycnal) component.

Applying a directional retention factor \(c\in[0,1]\) to the density-modifying subspace produces the filtered increment
\begin{equation}
 z_c=(I-E)z+cEz,\qquad \|z_c\|_w^2=\|z\|_w^2-(1-c^2)\|Ez\|_w^2.
 \label{eq:app-theory-norm}
\end{equation}
For DND (\(c=3/4\)), Eq.~\eqref{eq:app-theory-norm} reduces the squared volume-weighted norm of the density-modifying component \(Ez\) by \(1-(3/4)^2=7/16=43.75\%\), while leaving the density-compensated component \((I-E)z\) and the layerwise heat and salt budgets unchanged.

\subsection{Numerical implementation and norm-matched uniform control}
\label{app:theory-implementation}
\label{app:theory-local}

At each correction step, the layer-mean-preserving tracer increment is computed in 64-bit precision and normalized by \((s_T,s_S)\). Analytical partial derivatives \((\partial_T\rho,\partial_S\rho)\) of the POM equation of state (Eq.~\eqref{eq:app-eos}) evaluated at the pre-correction state define the unit normal \(n_i=g_i/\max(\|g_i\|_2,10^{-30})\). Within each active layer, the \(2\times2\) Gram matrix \(M=\sum_iw_in_in_i^\top=\sum_{j=1}^2\lambda_ju_ju_j^\top\) is inverted via its truncated eigendecomposition,
\begin{equation}
 M_\tau^+=\sum_{\lambda_j>\tau\lambda_{\max}}\max(\lambda_j,10^{-30})^{-1}u_ju_j^\top,\qquad \tau=10^{-10}.
\end{equation}
After forming \(z_c=(I-E)z+cEz\) with \(c=3/4\) and applying the paired \(L_\infty\) limiter (\(z_{\max}=5\), Appendix~\ref{app:training-correction}), the physical increments are added identically to \((T^n,T^b)\) and \((S^n,S^b)\), followed by a full nonlinear equation-of-state density update.

For the norm-matched Uniform control evaluated in Section~\ref{sec:analysis-uniform}, both tracer increments are scaled isotropically by \(z_\kappa^{\mathrm{uni}}=\kappa z\) with a fixed scalar factor \(\kappa\) calibrated once on cached Undamped development endpoints (days 1--10 from initialization days 8000, 8160, and 8320) to match the mean volume-weighted standardized tracer increment norm of DND: \(\kappa=0.9047\) for Implicit \(1\,\mathrm h\), \(0.9064\) for Implicit \(1\,\mathrm{day}\), \(0.9313\) for Explicit \(1\,\mathrm h\), and \(0.9038\) for Explicit \(1\,\mathrm{day}\). In the 72-hour single-perturbation propagation experiment (Section~\ref{sec:analysis-intervention}), the 36 pre-correction states are sampled at forecast lead days 1, 10, and 30 from initialization origins 8000, 8160, and 8320 across all four Posttrain configurations (\(3\times 3\times 4 = 36\) states).

\section{Evaluation Protocol and Physical Baselines}
\label{app:evaluation}

\subsection{Error metrics}
\label{app:evaluation-metrics}
\label{app:evaluation-weights}

For forecast initialization origin \(r\in\{1,\dots,16\}\) and lead day \(d\in\{1,\dots,100\}\), three-dimensional prognostic fields are weighted by cell volume \(w^{\mathrm{ref}}_{ijk}=A_{ij}m^f_{ij}(h_{ij}+\eta^{\mathrm{ref}}_{\mathrm B,ij})\Delta\sigma_k\), using the fine-grid reference body elevation of that day and the variable's native staggered wet mask; sea-surface height uses the horizontal cell area \(A_{ij}m_{ij}\). The spatial mean squared error \(M_f(r,d)\) is evaluated across the 15 active vertical layers, and any forecast lead window \([a,b]\) of length \(L=b-a+1\) days is summarized by the pooled root-mean-square error and relative percentage change:
\begin{equation}
 E_f(a,b)=\Bigl[\frac{1}{16L}\sum_{r=1}^{16}\sum_{d=a}^{b}M_f(r,d)\Bigr]^{1/2},\qquad
 \Delta_f^{\mathrm{ref}}=100\Bigl[\frac{E_f}{E_f^{\mathrm{ref}}}-1\Bigr].
 \label{eq:method-eval-rmse}
\end{equation}
All reductions are accumulated in 64-bit floating-point precision, whereas coarse-grid physical and hybrid forecast integrations execute at \(\Delta t=600\,\mathrm s\) in 32-bit floating-point precision. Training objectives and layer-budget constraints use the predicted model geometry, whereas forecast evaluation uses the fine-grid reference geometry common to all methods at each origin and lead day.

\subsection{Coarse-grid physical parameterizations}
\label{app:evaluation-baselines}

We compare three classical subgrid momentum and tracer closures on the coarse-grid POM dynamical frameworks:
\begin{itemize}
\item \textbf{Smagorinsky}: The standard second-order harmonic Smagorinsky horizontal viscosity \(A_0=c_m\,\Delta x\Delta y\,S_h\) and harmonic tracer diffusivity \(A_H=c_c\,\Delta x\Delta y\,S_h\), where \(S_h\) is the resolved horizontal deformation rate. The reference coarse-grid baseline uses \(c_m=0.05\) and \(c_c=0.01\) (\(\mathrm{tprni}=c_c/c_m=0.20\)), and we additionally sweep independent \((c_m,c_c)\) scaling pairs across both dynamical frameworks.
\item \textbf{Biharmonic}: A scale-selective fourth-order horizontal momentum dissipation closure that replaces harmonic momentum viscosity with a symmetric fourth-order dissipative update at each internal step,
\begin{equation}
 \boldsymbol v^+-\boldsymbol v^-=-\Delta t\,L^{\dagger_W}\kappa_4L\,\tfrac12(\boldsymbol v^++\boldsymbol v^-),\qquad
 \kappa_4=\gamma\frac{c_m}{0.05}\frac{A_0}{4(\Delta x^{-2}+\Delta y^{-2})},
 \label{eq:design-physical-dissipation}
\end{equation}
where \(L\) is the discrete horizontal stress operator at unit viscosity, \(W\) is the diagonal matrix of wet-face control volumes, and \(L^{\dagger_W}=W^{-1}L^\top W\) is its volume-weighted adjoint. The implicit midpoint update yields an exact single-step kinetic-energy dissipation rate of \(\Delta t\langle L\bar{\boldsymbol v},\kappa_4L\bar{\boldsymbol v}\rangle_W\) with \(\gamma=2\).
\item \textbf{Backscatter}: Fourth-order biharmonic viscosity coupled with a kinetic-energy backscatter parameterization. A fraction \(\beta=0.5\) of the biharmonic kinetic-energy loss is transferred to column subgrid energy reservoirs in regions where the flow is geostrophically balanced and the first baroclinic Rossby radius of deformation is resolved, and is reinjected into the resolved flow via anti-viscous harmonic forcing while accounting for the exact finite-step energy exchange (\(\gamma=2\)). Across both dynamical frameworks, Biharmonic and Backscatter yield larger 100-day RMSE than harmonic Smagorinsky across all five prognostic variables.
\end{itemize}

\subsection{Spatial tracer error decomposition and density}
\label{app:evaluation-tracers}

\begin{figure}[tbp]
\centering
\includegraphics[width=0.62\textwidth]{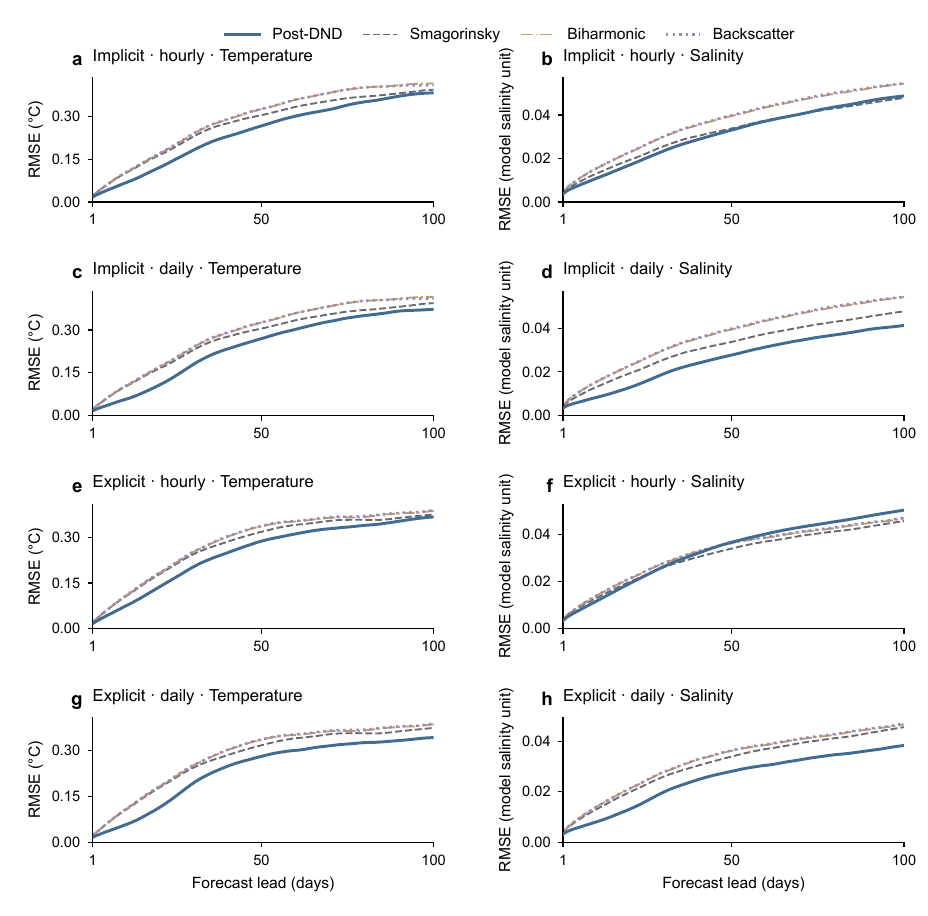}
\caption{\textbf{Tracer error evolution over 100 days.} Daily potential temperature and salinity RMSE of the Posttrain DND hybrid forecasts and the coarse-grid Smagorinsky baselines, pooled across the 16 evaluation origins at each lead day.}
\label{fig:error-growth}
\end{figure}

With normalized reference volume weights \(W_{ik}\) (\(\sum_{i,k}W_{ik}=1\)), layer weights \(W_k=\sum_iW_{ik}\), layer-mean error \(\mu_k=W_k^{-1}\sum_iW_{ik}e_{ik}\), and global domain-mean error \(\mu=\sum_kW_k\mu_k\), the three-dimensional tracer MSE decomposes orthogonally into domain-mean, vertical profile-shape, and horizontal anomaly components:
\begin{equation}
 \sum_{i,k}W_{ik}e_{ik}^2=\mu^2+\sum_kW_k(\mu_k-\mu)^2+\sum_{i,k}W_{ik}(e_{ik}-\mu_k)^2 .
 \label{eq:method-tracer-decomposition}
\end{equation}
In-situ density errors are evaluated in \(\mathrm{kg\,m^{-3}}\) from the nonlinear POM equation of state (Eq.~\eqref{eq:app-eos}). Figure~\ref{fig:error-growth} shows the 100-day daily RMSE evolution of temperature and salinity across all four hybrid configurations and their coarse-grid Smagorinsky baselines.

\subsection{Horizontal velocity spectra}
\label{app:evaluation-spectra}

Each horizontal velocity component is analyzed on its native staggered C-grid. Within each active layer \(k\), the velocity field is masked, multiplied by a two-dimensional Hann window \(q_{ij}\), demeaned with the same window weights, normalized by \(\bigl[(N_xN_y)^{-1}\sum_{ij} q_{ij}^2\bigr]^{1/2}\), and transformed via forward-normalized two-dimensional fast Fourier transforms. Isotropic radial wavenumber bins span \(0\) to \(\sqrt{1/2}\) cycles per grid interval. Layer kinetic energies \(E_{kb}=\tfrac12\sum_{\mathcal B_b}(|\hat u_k|^2+|\hat v_k|^2)\) are weighted by layer thickness \(\Delta\sigma_k\). Prediction, reference, error, and cross-spectral kinetic energies \(E^p,E^r,E^e,C\) are averaged across the 16 evaluation origins and lead days within each forecast window before forming the energy ratio \(R=E^p/E^r\), normalized spectral error \(Q=E^e/E^r\), and phase alignment \(A=C/\sqrt{E^pE^r}\), which satisfy the exact identity \(Q=R+1-2A\sqrt R\). Spectral bands are partitioned into large scales (\(k\le0.125\) cycles per grid interval), intermediate scales (\(0.125<k\le0.25\)), near-grid scales (\(0.25<k\le0.5\)), and corner scales (\(k>0.5\)).
\FloatBarrier

\section{Direct Neural Forecasters}
\label{app:ai-comparison}

\subsection{Full-state emulation task and architectures}

The direct autoregressive neural forecasters are trained and evaluated on the semi-implicit reference dataset in the wind-driven double-gyre ocean basin using the same ensemble member and temporal split as NeuralPOM. At each daily step, each direct model takes as input and predicts the complete 313-channel model state: 18 three-dimensional variables across all 16 vertical levels (current and backward horizontal velocities and tracers, turbulent kinetic energy and mixing length variables, vertical momentum and tracer diffusivities, horizontal viscosity, in-situ density, and transformed vertical velocity \(\Omega\)) together with 25 two-dimensional variables (free-surface elevations, depth-averaged velocities, interval transport accumulators, surface wind forcing, and bottom stress). Each channel is standardized using frozen training-set statistics; land and bottom cells are masked, and the model clock advances by one day per autoregressive step without invoking the primitive-equation solver.

We evaluate five spatiotemporal neural architectures:
\begin{itemize}
\item \textbf{ResUNet}: The residual U-Net architecture of the hybrid corrector widened to 313 input and 313 output channels.
\item \textbf{Triton}: The multi-scale spatiotemporal hierarchy of \cite{Wu2025TritonV1}, configured with spatial hidden width 256, temporal hidden width 512, four spatial blocks, and eight temporal blocks.
\item \textbf{SimVP}: The Inception-based spatiotemporal translator of \cite{Gao2022SimVP} with spatial width 256 and temporal width 512.
\item \textbf{ConvLSTM}: A two-layer convolutional LSTM network \cite{Shi2015ConvLSTM} with 256 hidden channels per layer and \(3\times3\) kernels.
\item \textbf{PastNet}: The hybrid spectral-convolution and discrete vector-quantized codebook architecture of \cite{Wu2023PastNet} with a 512-entry codebook.
\end{itemize}
Horizontal grids are padded to \(92\times84\) (Triton, SimVP) or \(96\times96\) (PastNet) and cropped back to the \(90\times82\) domain after each forward pass.

\subsection{Two-stage training protocol and window evaluation}

All five direct neural architectures are trained using the channel-averaged, area- and thickness-weighted mean squared error across all 313 standardized state channels:
\begin{equation}
 \ell_{\mathrm{AI}}=\frac1{313}\sum_{c=1}^{313}\frac{\sum_iw_{ci}(\hat a_{ci}-a_{ci})^2}{\sum_iw_{ci}} .
 \label{eq:ai-full-state-loss}
\end{equation}
During single-step pretraining (Pretrain), each network is trained on one-day state transitions sampled every six days across the seven 6000-day training members using the same optimizer and cosine learning-rate schedule as the hybrid Pretrain stage. During closed-loop rollout training (Posttrain), initialized from the Pretrain checkpoint, each network is trained on 10-day autoregressive rollouts sampled every 60 days, performing one optimizer update per rollout day with the predicted state detached between consecutive daily steps. In both stages, training is terminated via early stopping based on validation skill.

Whereas Table~\ref{tab:ai-main} in Section~\ref{sec:results-ai} reports the full 100-day pooled RMSE across all 16 evaluation origins, Table~\ref{tab:ai-windows} partitions performance into early (days 1--50) and late (days 51--100) forecast windows for all ten direct neural checkpoints, the coarse-grid Smagorinsky baseline, and the daily NeuralPOM hybrid forecast (DND).

\begin{table}[tbp]
\centering\small
\caption{\textbf{Early- and late-window forecast RMSE (days 1--50 and days 51--100).} Shading and boldface denote the lowest RMSE in each column within each forecast window; underlining denotes the second lowest (see Table~\ref{tab:ai-main} for days 1--100). Horizontal velocities \(u,v\) in \(\mathrm{m\,s^{-1}}\), sea-surface height (SSH) in \(\mathrm{m}\), potential temperature \(T\) in \(^\circ\mathrm{C}\), and salinity \(S\) in model salinity units.}
\label{tab:ai-windows}
\setlength{\tabcolsep}{4pt}
\makeatletter\let\AIprimitiveinput\@@input\makeatother
\begin{tabular}{lrrrrr}
\toprule
Configuration & $u$ $\downarrow$ & $v$ $\downarrow$ & SSH $\downarrow$ & $T$ $\downarrow$ & $S$ $\downarrow$\\
\midrule
\multicolumn{6}{l}{\emph{Days 1--50}}\\
\AIprimitiveinput data/ai_baselines/table_1_50_rows.tex
\midrule
\multicolumn{6}{l}{\emph{Days 51--100}}\\
\AIprimitiveinput data/ai_baselines/table_51_100_rows.tex
\bottomrule
\end{tabular}
\end{table}
\FloatBarrier

\section{Validation of the differentiable dynamical frameworks}
\label{app:validation}
\label{app:evaluation-cases}
\label{app:physics-adaptations}

Both differentiable PyTorch dynamical frameworks were validated in 64-bit floating point against the double-precision Fortran Princeton Ocean Model across four benchmark configurations spanning idealized basin, coastal, topographic, and realistic embayment regimes with identical initial states, bathymetry, forcing, and time steps:
\begin{enumerate}
\item \textbf{Double gyre}: A 24-hour spin-up of the wind-driven double-gyre ocean basin on a closed \(90\times82\) horizontal grid with 15 active \(\sigma\)-layers, zero normal boundary velocities, internal step \(\Delta t=900\,\mathrm s\), external substep \(\delta t=30\,\mathrm s\), and \(\mu=0.10\) (whereas the long double-gyre forecast integrations in the main text use \(\Delta t=600\,\mathrm s\), \(\delta t=20\,\mathrm s\), \((\alpha,\mu)=(0.225,0.10)\) for Explicit, and \((0.333,0.15)\) for Implicit).
\item \textbf{Coastal tide}: A 15-day barotropic tidal integration on a \(161\times101\) grid with four active layers, \(\Delta t=10\,\mathrm s\), \(\delta t=1\,\mathrm s\), fixed tracers, logarithmic bottom drag (Eq.~\eqref{eq:app-coastal-drag} with \(z_{\mathrm{sh}}=0.01\,\mathrm m\) and \(z_0=0.005\,\mathrm m\)), an eastern Flather radiation condition \(U_E=U_{\mathrm{adj}}+\sqrt{g/H_{\mathrm{adj}}}(\eta_{\mathrm{adj}}-\eta_E)\), and western \(\mathrm M_2\) tidal forcing \(\eta_{\mathrm{tide},j}(t_h)=a_j\cos(\omega t_h-\phi_j)\) enforced via Dirichlet boundary lifting in the Implicit solver.
\item \textbf{Stratified seamount}: A 10-day integration initialized from rest over an isolated seamount on a \(65\times49\) grid with 15 active layers, \(\Delta t=150\,\mathrm s\), \(\delta t=5\,\mathrm s\), \(r_{\mathrm{mean}}=0\), open eastern and western radiation boundaries \(U_{W,E}=\mp\sqrt{g/h}(\eta_{W,E}-\eta_{W,E}^{\mathrm{ext}})\) incorporated into \(A_\eta\) as diagonal Robin terms, and conservative body-elevation updates driven by \(\overline{\mathcal Q}\).
\item \textbf{Jiaozhou Bay}: A 15-day realistic-topography tidal integration of Jiaozhou Bay on a \(312\times213\) horizontal grid (\(\Delta x\approx141.7\,\mathrm m\), \(\Delta y\approx185.0\,\mathrm m\), \(30{,}159\) interior wet water columns with \(2.40\,\mathrm m\le h\le60.03\,\mathrm m\)) and 16 active \(\sigma\)-layers, with \(\Delta t=10\,\mathrm s\), \(\delta t=1\,\mathrm s\), fixed tracers, depth-gated logarithmic bottom drag (\(z_{\mathrm{sh}}=0.01\,\mathrm m\), \(z_0=0.005\,\mathrm m\) evaluated where \(H\ge h_c=0.05\,\mathrm m\)), gravity-wave and Orlanski velocity radiation boundaries, and \(\mathrm M_2\) tidal forcing across 389 active wet open-boundary nodes along the southern and eastern boundaries enforced via Dirichlet boundary lifting in the Implicit solver.
\end{enumerate}
Across all four validation benchmarks, the Implicit solver uses the conservative transport corrector (Eq.~\eqref{eq:app-conservative-corrector}), area-weighted PCG inner products, and a relative residual tolerance of \(10^{-11}\).

\begin{figure}[!htbp]
\centering
\includegraphics[width=0.82\textwidth]{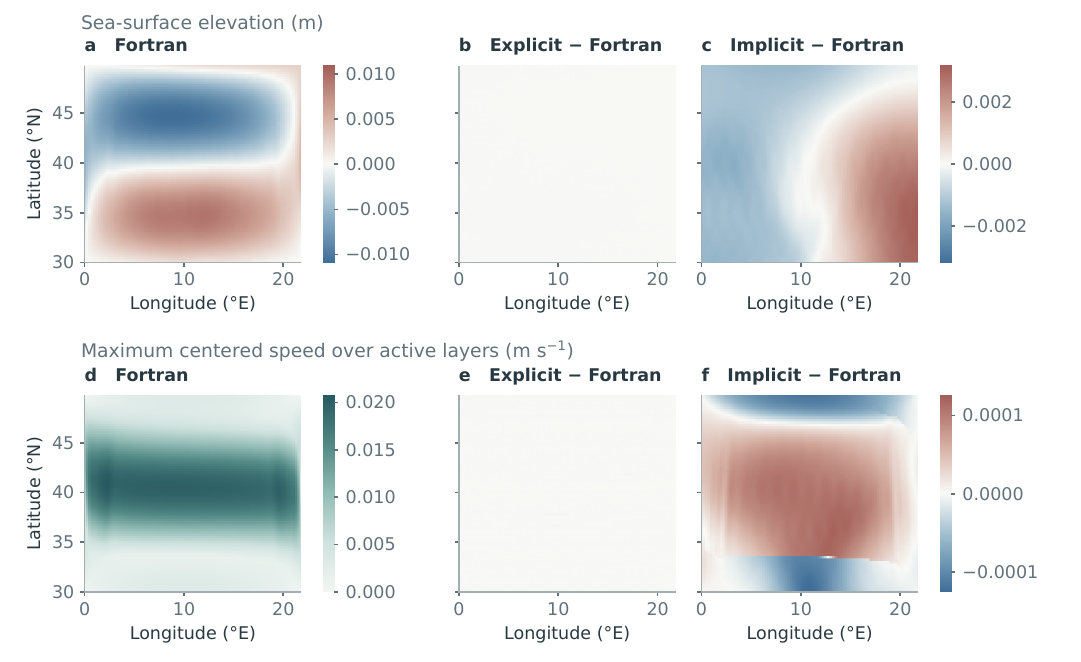}
\caption{\textbf{Double-gyre validation benchmark at 24 hours.} Fortran POM reference fields (left) and pointwise differences for the Explicit (middle) and Implicit (right) differentiable PyTorch frameworks, shown for free-surface elevation (top) and column-maximum horizontal speed (bottom).}
\label{fig:dg-fields}
\end{figure}

\begin{figure}[!htbp]
\centering
\includegraphics[width=0.96\textwidth]{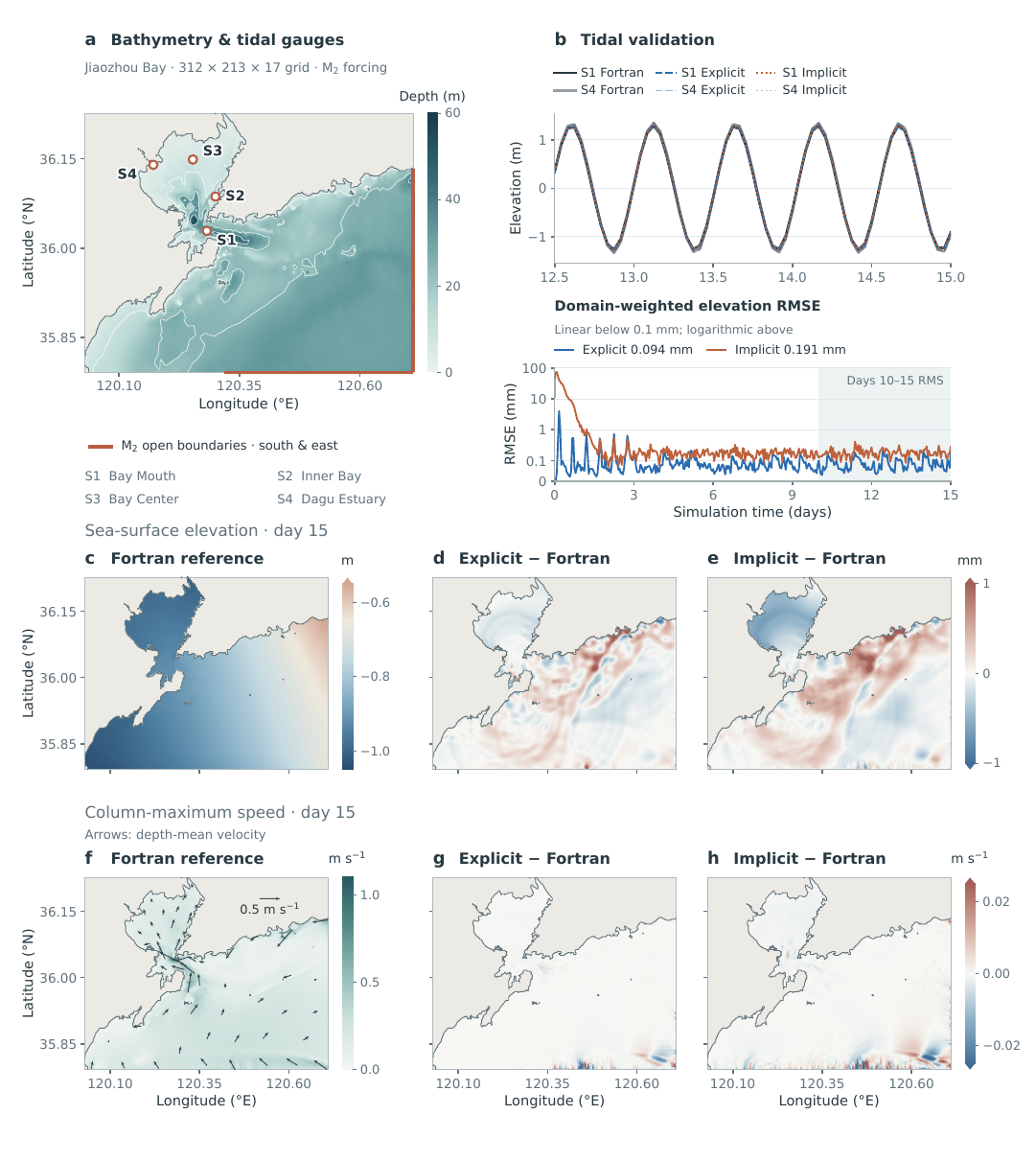}
\caption{\textbf{Realistic-topography Jiaozhou Bay validation benchmark over a 15-day (\boldmath\(360\,\mathrm h\)) \(\mathrm M_2\) tidal integration.} (\textbf{a}) Coastal bathymetry \(h\), southern and eastern \(\mathrm M_2\) open boundaries, and interior tidal gauge locations (\(\mathrm{S1}\) Bay Mouth, \(\mathrm{S2}\) Inner Bay, \(\mathrm{S3}\) Bay Center, \(\mathrm{S4}\) Dagu Estuary). (\textbf{b}) Equilibrium tidal gauge elevation series over days 12.5--15 (top) and 15-day domain-weighted free-surface elevation RMSE time series against the double-precision Fortran POM reference (bottom). (\textbf{c--e}) Day-15 free-surface elevation \(\eta\) for the Fortran reference (\(\mathrm m\)) and pointwise differences (\(\mathrm{mm}\)) for the Explicit and Implicit differentiable PyTorch frameworks. (\textbf{f--h}) Day-15 column-maximum horizontal speed (\(\mathrm{m\,s^{-1}}\)) with depth-averaged current vectors overlaid on the Fortran reference, alongside pointwise speed differences for the Explicit and Implicit frameworks.}
\label{fig:jzb-fields}
\end{figure}

The Explicit framework reproduces the Fortran reference to numerical round-off: after 24 hours in the double-gyre benchmark, its velocity RMSEs are \(1.78\times10^{-9}\) and \(1.64\times10^{-9}\,\mathrm{m\,s^{-1}}\), and its sea-surface elevation RMSE is \(8.51\times10^{-10}\,\mathrm m\) (Table~\ref{tab:case-errors} and Fig.~\ref{fig:dg-fields}). The Implicit framework replaces external mode subcycling with a semi-implicit elliptic solve at the baroclinic step; its 24-hour double-gyre elevation difference of \(1.42\,\mathrm{mm}\) decreases to \(0.46\,\mathrm{mm}\) when \(\Delta t\) is reduced from \(900\,\mathrm s\) to \(100\,\mathrm s\), confirming temporal consistency between the two free-surface discretizations. In the coastal tidal benchmark, the root-mean-square elevation difference over hours 240--360 is \(1.9\,\mu\mathrm m\) for Explicit and \(0.50\,\mathrm{mm}\) for Implicit, with station \(\mathrm M_2\) tidal amplitudes and phases agreeing with Fortran to within \(0.46\,\mathrm{mm}\) and \(0.042^\circ\). In the resting stratified seamount benchmark, both PyTorch frameworks and Fortran POM develop the same terrain-following pressure-gradient error current of approximately \(0.30\,\mathrm{m\,s^{-1}}\) after 10 days \cite{ShchepetkinMcWilliams2003Pressure}, with inter-model velocity differences below \(10^{-4}\,\mathrm{m\,s^{-1}}\). In the realistic-topography Jiaozhou Bay benchmark, where an un-ramped cold-start \(\mathrm M_2\) tide propagates across complex coastal geometry and shallow shoals, the Explicit framework matches Fortran to machine precision over the initial steps (\(\sim10^{-16}\,\mathrm m\)) and maintains a root-mean-square elevation difference of \(0.094\,\mathrm{mm}\) over days 10--15 (\(0.19\,\mathrm{mm}\) at day 15), while the Implicit framework damps the initial six-hour cold-start gravity-wave transient at \(\Delta t=10\,\mathrm s\) before converging onto the forced tidal equilibrium with a root-mean-square elevation difference of \(0.19\,\mathrm{mm}\) over days 10--15 (\(0.28\,\mathrm{mm}\) at day 15) and interior station \(\mathrm M_2\) amplitude and phase differences within \(0.12\,\mathrm{mm}\) and \(0.010^\circ\) across both frameworks (Fig.~\ref{fig:jzb-fields}).

\begin{table}[htbp]
\centering\small
\caption{\textbf{Root-mean-square differences against the double-precision Fortran POM reference at the end of each validation benchmark.} Velocities \(u,v\) in \(\mathrm{m\,s^{-1}}\), potential temperature \(T\) in \(^\circ\mathrm C\), salinity \(S\) in model salinity units, and free-surface elevation \(\eta\) in \(\mathrm m\).}
\label{tab:case-errors}
\begin{tabular}{@{}llrrrrr@{}}
\toprule
Benchmark & Framework & $u$ & $v$ & $T$ & $S$ & $\eta$\\
\midrule
Double gyre (24 h) & Explicit & 1.780e-09 & 1.638e-09 & 3.502e-09 & 1.981e-09 & 8.510e-10\\
Double gyre (24 h) & Implicit & 7.357e-05 & 3.563e-05 & 2.303e-06 & 1.438e-05 & 1.416e-03\\
Coastal tide (15 d) & Explicit & 1.349e-06 & 7.935e-07 & 0 & 0 & 1.224e-06\\
Coastal tide (15 d) & Implicit & 7.175e-04 & 1.858e-04 & 0 & 0 & 5.466e-04\\
Seamount (10 d) & Explicit & 6.298e-05 & 2.382e-05 & 1.764e-04 & 9.766e-08 & 4.439e-05\\
Seamount (10 d) & Implicit & 9.304e-05 & 4.911e-05 & 1.973e-04 & 9.540e-08 & 3.002e-05\\
Jiaozhou Bay (15 d) & Explicit & 2.916e-03 & 5.724e-03 & 0 & 0 & 1.866e-04\\
Jiaozhou Bay (15 d) & Implicit & 5.371e-03 & 7.468e-03 & 0 & 0 & 2.840e-04\\
\bottomrule
\end{tabular}
\end{table}

\end{document}